\documentclass[11pt]{article}
\usepackage{amsmath, amssymb, amsthm, epsfig}
\newcommand{\llq}{\leq}

\newtheorem{theorem}{Theorem}[section]

\newtheorem{remark}[theorem]{Remark}
\newtheorem{lemma}[theorem]{Lemma}
\newtheorem{corollary}[theorem]{Corollary}

\title{Eigenvalues of Hermite and Laguerre ensembles: Large Beta Asymptotics} 
\author{Ioana Dumitriu and Alan Edelman}

\begin{document}

\maketitle

\abstract{In this paper we examine the zero and first order eigenvalue fluctuations for the $\beta$-Hermite and $\beta$-Laguerre ensembles, using the matrix models we described in \cite{dumitriu02}, in the limit as $\beta \rightarrow \infty$. We find that the fluctuations are described by Gaussians of variance $O(1/\beta)$, centered at the roots of a corresponding Hermite (Laguerre) polynomial. We also show that the approximation is very good, even for small values of $\beta$, by plotting exact level densities versus sum of Gaussians approximations.}

\section{Introduction}

    This paper provides insight into the shape of random
matrix laws such  as the finite semi-circle law, the
finite quarter-circle law and its generalization.
    We begin with a simple example.  Suppose $A$ is a random
$k \times k$ complex matrix with real and imaginary parts all i.i.d. standard
normals.  Let $S=(A+A^H)/2$ be the Hermitian part of $A$.
The matrix $S$ has a distribution commonly known as the
Gaussian Unitary Ensemble; this matrix distribution and the joint distribution
of its (real) eigenvalues have been well studied. For 
a good reference on the subject, see Mehta \cite{mehta_book}.

    We draw below histograms of normalized eigenvalues taken from this
distribution, the known theoretical distribution (see \cite[page 93]{mehta_book}), and the semicircle limit corresponding to $k \rightarrow \infty$. For the histograms, we have chosen $40,000$ samples from the GUE with $k=4$ and $k=6$.

\begin{figure}[ht!]
\hspace{.75cm} \parbox[b]{5cm}{\epsfig{figure =  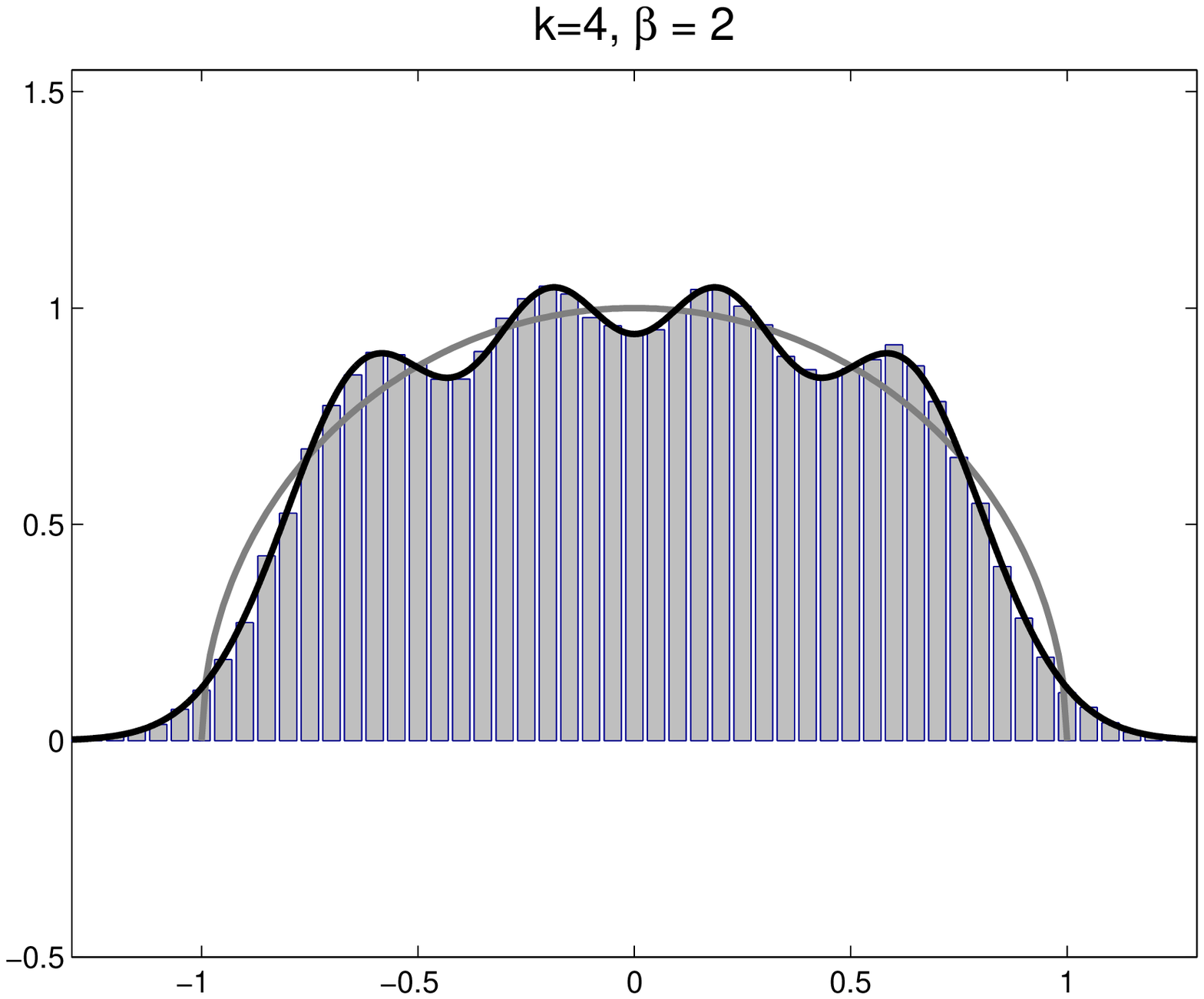, height = 4.75cm}} \hspace{1.75cm}
\parbox[b]{5cm}{\epsfig{figure = 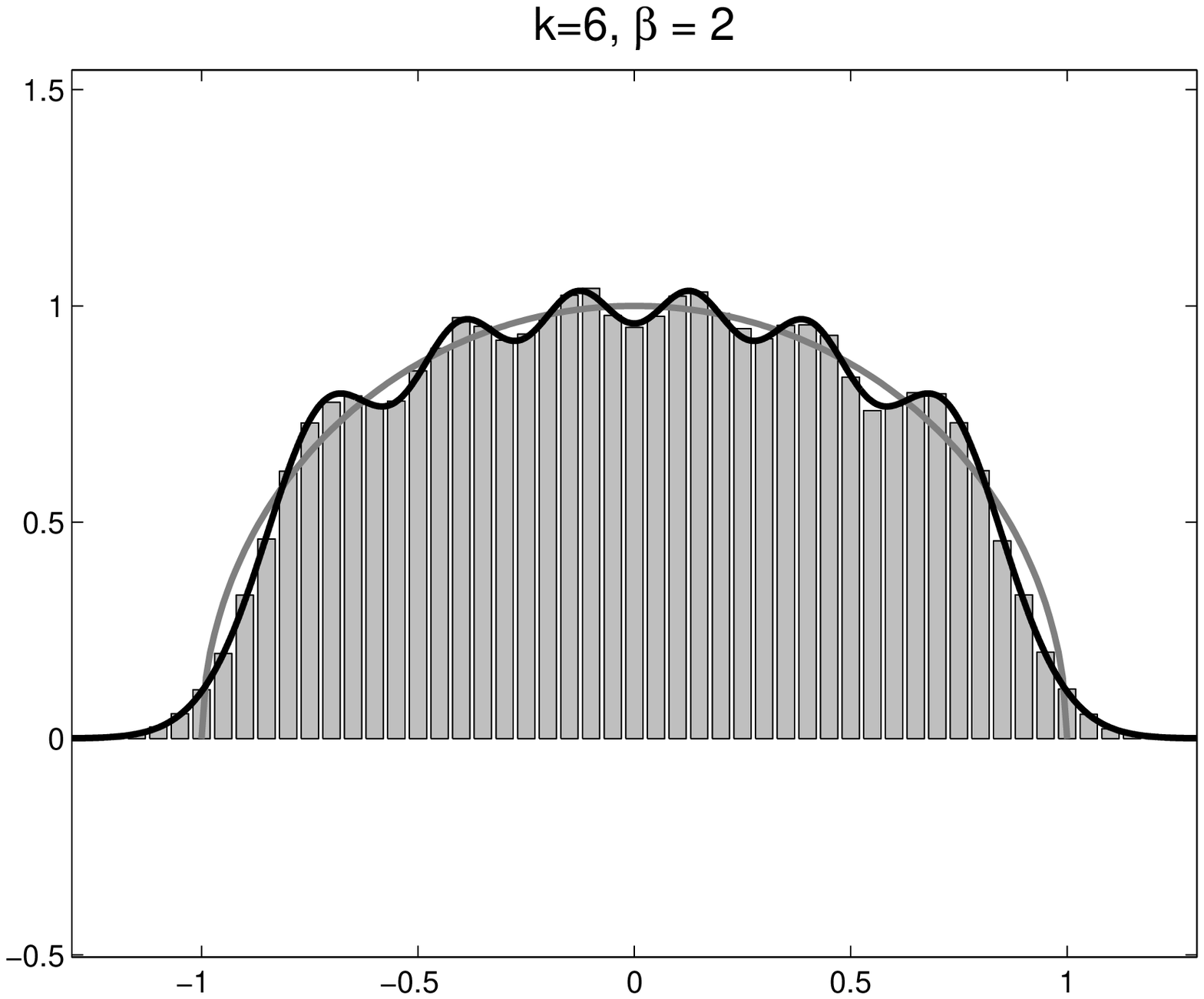, height = 4.75cm} }
\caption{Histograms of eigenvalues, finite ($k=4,6$) exact level densities, and semicircle law ($k \rightarrow \infty$) for matrices of the GUE ensemble} \label{combined_semi_4_6}
\end{figure}


Notice, for $k$ finite, the $k$ ``bumps'' in the distribution that wiggle above
and below the semi-circle.  A natural question to many engineers, physicists,
mathematicians, and other scientists who have seen these pictures is
whether they can be well approximated by the sum of $k$ appropriately
chosen Gaussians.  (Of course when $k=1$, this is exactly true.) 
The answer, as proved in this paper, is yes. We give
a sum of Gaussians approximation that is asymptotically
correct for the $\beta \rightarrow \infty$ limit but
useful even for small values of $\beta$.

For those well versed in random matrix theory, the GUE is the
$\beta=2$ case of a Hermite matrix ensemble \cite{mehta_book}.
Had we started with $A$ real (quaternion), we would have
the Gaussian Orthogonal (Symplectic) Ensemble corresponding
to $\beta=1$ ($\beta=4$).

The joint eigenvalue density $f_{\beta}^{H}(\lambda_1, \ldots, \lambda_k)$ defined on $\mathbb{R}^k$ for the $k$ eigenvalues
for an arbitrary $\beta>0$ is given in the formula below. 
\begin{eqnarray} \label{herm_ensembles}
f_{\beta}^{H}(\lambda_1, \ldots, \lambda_k) = (2 \pi)^{-k/2} \prod_{j=1}^{k} \frac{\Gamma (1 + \frac{\beta}{2})}{\Gamma ( 1+ j \frac{\beta}{2})} ~~\prod\limits_{1 \leq i<j \leq k} |\lambda_i - \lambda_j|^{\beta} ~e^{-\sum\limits_{i=1}^k \lambda_i^2/2}~.
\end{eqnarray}
Note the ``repulsion'' factor $\Delta(\Lambda) \equiv \Delta(\lambda_1, \ldots, \lambda_k) \equiv \prod\limits_{1 \leq i<j \leq k} |\lambda_i - \lambda_j|$.

Similarly, for the $k \times k$ Laguerre ensembles of statistics (Wishart matrix theory), the joint eigenvalues density $F_{\beta, a}^{L}$ is defined on $[0, \infty)^k$ for arbitrary $\beta$ and parameter $a>(k-1) \beta/2$, and is given below; once again note the repulsion factor $\Delta(\Lambda)$:
\begin{eqnarray} \label{lag_ensembles}
f_{\beta, a}^{L}(\lambda_1, \ldots, \lambda_k) = c_{\beta, a}^{L} ~~\prod\limits_{1 \leq i<j \leq k} |\lambda_i - \lambda_j|^{\beta} ~ \prod_{i=1}^k \lambda_i^{a - (k-1) \beta/2 - 1} ~e^{-\sum\limits_{i=1}^k \lambda_i/2}~,
\end{eqnarray}
where 
\[
c_{\beta, a}  = 2^{-ka} \prod_{j=1}^k \frac{\Gamma(1+ \frac{\beta}{2})}{\Gamma(1+ j \frac{\beta}{2}) \Gamma(a-(k-j) \frac{\beta}{2})}~.
\]

In \cite{dumitriu02}, we have found (real) tridiagonal matrix models whose eigenvalue distributions are given by \eqref{herm_ensembles} and \eqref{lag_ensembles}; we depict the distributions in Table \ref{table}. Note that the variables have either standard normal distribution or a $\chi$ distribution (sometimes scaled by $\sqrt{2}$).

\begin{table}[ht] \begin{center} \begin{tabular}{|l||l|} \hline & \\ $\begin{array}{c}
\mbox{\textbf{Hermite} matrix} \\ n \in \mathbb{N} \end{array}$ & {\huge $A_{\beta}$} $
\sim \frac{1}{\sqrt{2}} \left( \begin{array}{ccccc} N(0, 2) & \chi_{(n-1) \beta} & & & \\
\chi_{(n-1) \beta} & N(0, 2)  & \chi_{(n-2) \beta} & & \\ & \ddots & \ddots & \ddots & \\
& & \chi_{2\beta} & N(0,2) & \chi_{\beta} \\ & & & \chi_{\beta} & N(0,2) \end{array}
\right)$ \\ & \\ \hline & \\ \mbox{\textbf{Laguerre} matrix} & {\huge $L_{\beta}$} $=
B_{\beta}^{} B_{\beta}^{T}$, where \\ $\begin{array}{c} m \in \mathbb{N} \\ a \in
\mathbb{R} \\ a>\frac{\beta}{2}(m-1) \end{array}$ & \hspace{1cm} $B_{\beta} \sim \left(
\begin{array} {cccc} \chi_{2a} & & & \\ \chi_{\beta (m-1)} & \chi_{2a-\beta} & & \\ &
\ddots & \ddots & \\ & & \chi_{\beta} & \chi_{2a-\beta(m-1)} \end{array} \right)$ \\ & \\
\hline \end{tabular} \end{center} \caption{Tridiagonal matrix models for the $\beta$-Hermite and $\beta$-Laguerre ensembles with any $\beta>0$.} \label{table}
\end{table}
                                                                                  
For generating efficiently eigenvalues for the $\beta$-ensemble distributions, we recommend using the tridiagonal/bidiagonal model above. 

The marginal density of a single eigenvalue (known as the level density) 
can be computed
in the case of the Hermite ensembles for $\beta$ an even integer \cite{Forrester_poly} using the MOPS software \cite{dumitriu04a}.
For general $\beta$ perhaps a little more research is needed, but
it is likely that computational techniques are not far out of reach.

For fixed $k$ and general $\beta$ one finds that for $\beta$ getting
larger, the bumps of Figure \ref{combined_semi_4_6} get ``bumpier''.  To be precise,
for the Hermite ensembles, we prove here that at $\beta=\infty$ the  bumps become delta functions 
at the roots of the $k$th Hermite polynomial, while for $\beta$ large, 
the bumps behave like Gaussians centered at these roots with 
variance $O(1/\beta)$.

The model of $\beta$ as an inverse temperature is apparent from 
\eqref{herm_ensembles}.  As $\beta $ goes to $0$, the strength of the repulsion factor $\Delta(\Lambda)$ decreases until annihilation; the interdependence among eigenvalues disappears, and the randomness increases (each eigenvalue behaves like an independent Gaussian). In the frozen state ($\beta = \infty$), we can imagine the $k$ eigenvalues fixed at the roots of the Hermite polynomial. Warming the system a little ($\beta$ very large but not infinite) gives the particles a little energy, and the eigenvalues have Gaussian distribution to first order around the Hermite polynomial roots.

Similarly, in the Laguerre case, at $\beta =0$, the eigenvalues become i.i.d. variables with distribution $\chi_{2a}^2$. As $\beta$ grows the eigenvalues have Gaussian distribution to first order around the Laguerre polynomial roots, while at $\beta = \infty$ we reach the freezing point when the eigenvalues are fixed at those roots. 

In the following, we use first order eigenvalue perturbation theory
and the tridiagonal ensembles in \cite{dumitriu02} to rigorously
investigate this phenomenon mathematically obtaining precisely
the asymptotic variance along with the mean. 

These results draw a parallel to the Tracy-Widom laws \cite{tracy_widom_1_4, tracy_widom_largest} for the $\beta = 1,2,4$-Hermite ensembles, later extended to $\beta = 1,2$-Laguerre ensembles by Johansson \cite{johansson_wishart} and Johnstone \cite{Johnstone}. 

The Tracy-Widom laws compute the fluctuation in the distribution of the largest eigenvalue of a $\beta$-Hermite ensemble with $\beta = 1,2,4$, as $k \rightarrow \infty$, and obtain it in terms of the solution to a Painleve differential equation. From the semicircle law, we know that as $n \rightarrow \infty$, regardless of $\beta$, the largest eigenvalue (scaled by $\sqrt{2k\beta}$) goes to $1$. From Theorem \ref{beta_large_herm}, constrained to $i=1$, with the help of \cite{szego} and \cite[page 450]{Abr-Steg}, we obtain Corollary \ref{corr1}, which gives an intuition of how the $\beta = 1,2,4$ Tracy-Widom distributions evolve towards a normal distribution at $\beta = \infty$. 

The theoretical results of Section \ref{applications} are similar to the ``Central Limit Theorems'', i.e. the computation of the global fluctuations from the semicircle and semicircle-type laws done by Johansson in \cite{johansson_clt_herm} for Hermite-like ensembles of any $\beta$ and by Silverstein and Bai \cite{silverstein03} for a class of Laguerre-like ensembles with real or complex entries ($\beta = 1,2$). Roughly said, the eigenvalues can be thought of as fluctuating (like Gaussians) around the roots of the corresponding orthogonal polynomial as $\beta$ grows large; if one lets $n$ grow large, too, the global eigenvalue fluctuation becomes a Gaussian process. The larger $1/\beta$, the ``warmer'' it gets, and the larger the ``vibration''. The larger $\beta$, the ``cooler'' it gets, and the eigenvalues ``freeze'' into place.

At the end of Section \ref{applications} we perform computational experiments to see how
good the $\beta$ large approximation is even for relatively small
$k$ and $\beta$.

\section{Eigenvalue perturbation and $\chi$ asymptotics}

In this section we present two lemmas we need in the proofs of our main results (Theorems \ref{beta_large_herm} and \ref{beta_large_lag}). 

The first lemma involves perturbation theory; for a good reference on Perturbation theory and a more general form of the result below, see Demmel's book \cite[Section 4.3]{demmel_book}.

\begin{lemma} \label{perturb_lemma}Let $A$ and $B$ be $n \times n$ symmetric matrices, and let $\epsilon>0$. Assume $A$ has all distinct eigenvalues. Let $M = A+ \epsilon B + o(\epsilon)$, where by $o(\epsilon)$ we mean a matrix in which every entry goes to $0$ faster than $\epsilon$. Let $\lambda_i(X)$ denote the $i$th eigenvalue of $X$, for $1 \leq i \leq n$. Finally, let $Q$ be an eigenvector matrix for $A$. Then 
\[
\lim_{\epsilon \rightarrow 0} \frac{1}{\epsilon} (\lambda_i(M) - \lambda_i(A)) = Q(;, i)^{T} B Q(:, i)~,
\]
where, following MATLAB notation, $Q(:, i)$ represents the $i$th column of $Q$.
\end{lemma}

\begin{remark} Equivalently, for every $1 \leq i \leq n$, 
\[
\lambda_i(M) = \lambda_i(A)+ \epsilon Q(;, i)^{T} B Q(:, i) + o(\epsilon)~.
\]
\end{remark}

The second result is an approximation lemma for the $\chi_r$ distribution as $r$ grows large.

\begin{lemma} \label{approx_chi} Let $r>0$, and let $X$ be a variable with distribution $\chi_r$. Then as $r \rightarrow \infty$ the distribution of $X-\sqrt{r}$ converges to a normal distribution of mean $0$ and variance$~1/2$. 
\end{lemma}

\begin{proof}
We prove this lemma by looking at the density function of $\chi_{r}$ when $r \rightarrow \infty $. 

First we will show that $E[X] - \sqrt{r} \rightarrow 0$ as $r \rightarrow \infty$. Since
\begin{eqnarray} \label{what_I_need}
E[X] = \frac{2^{1- r/2}}{\Gamma \left( \frac{r}{2} \right )} \frac{\Gamma \left ( \frac{r+1}{2} \right)}{2^{1 - (r+1)/2}} = \sqrt{2} \frac{\Gamma \left ( \frac{r+1}{2} \right)}{\Gamma \left( \frac{r}{2} \right )}~.
\end{eqnarray}

Using the Stirling approximation formula 
\begin{eqnarray} \label{stirling}
\Gamma(z) & \sim & z^{z-1/2} e^{-z}  \sqrt{2\pi} \left(1+ \frac{1}{12z} +O(\frac{1}{z^2}) \right)~,
\end{eqnarray}
for $r$ large, (\ref{what_I_need}) becomes
\begin{eqnarray*}
E[X] & = & \sqrt{r} ~(1+O (r^{-1}))~.
\end{eqnarray*}

So as $r \rightarrow \infty$, the mean of $X - \sqrt{r}$ approaches $0$. Now let us examine the p.d.f.\ of $X - \sqrt{r}$. Denote by $Y = X - \sqrt{r}$; the p.d.f.\ of $Y$ is
\[
f(t) = \frac{2^{1-r/2}}{\Gamma \left(\frac{r}{2}\right)} ~(t + \sqrt{r})^{r-1} e^{-(t+\sqrt{r})^2/2}~.
\]

We examine this p.d.f.\ in a ``small'' neighborhood of $0$, such that $t = o(r^{1/2})$. Again, we use the Stirling  approximation (\ref{stirling}) for the Gamma function at infinity, to obtain
\begin{eqnarray*}
f(t) & = & \frac{2^{1- r/2} ~r^{r/2-1}~ \left(1+\frac{t}{r}\right)^{r-1} e^{-t^2/2 - \sqrt{r}t - r/2} }{\left(\frac{r}{2} -1\right)^{r/2-1} ~e^{-r/2+1} ~\sqrt{\pi r}} ~\left(1+O(r^{-1}) \right) ~~, \\
& =& \frac{1}{\sqrt{\pi}} ~\left(1+\frac{t}{\sqrt{r}}\right)^{r-1}~ e^{-t^2/2 - \sqrt{r} t} ~\left(1+O(r^{-1}) \right) ~,
\end{eqnarray*}
and by using the Taylor series for $(r-1)\ln(1+t/\sqrt{r})$ and the fact that $t = o(r^{1/2})$, we obtain
\begin{eqnarray} \label{key}
f(t) = \frac{1}{\sqrt{\pi}} e^{-t^2} ~\left(1+O\left(\frac{t}{\sqrt{r}} \right ) \right) ~.
\end{eqnarray}

Thus, on any fixed interval, the p.d.f.\ of $X - \sqrt{r}$ converges to the p.d.f.\ of a centered normal of variance $1/2$. This is enough to prove that the c.d.f. of $Y = X - \sqrt{r}$ converges to the c.d.f. of a centered normal of variance $1/2$.
\end{proof}

\section{$\beta$-Hermite: zero and first-order approximations} \label{eigs_herm}

Let $k$ be fixed, and let $h_1^{(k)}, \ldots, h_k^{(k)}$ be the roots of the $k$th univariate Hermite polynomial  $H_k$ (where the Hermite polynomial\index{Hermite polynomial}s $H_0(x), H_1(x), \ldots$ are orthonormal with respect to the weight $e^{-x^2}$ on $(-\infty, \infty)$). 

Let $A_{\beta}$ be a random matrix from the $\beta$-Hermite ensemble of size $k$, scaled by $1/\sqrt{2k\beta}$. For the remainder of this section, we think of $\beta$ as a parameter.


We state and prove the following theorem.

\begin{theorem} \label{beta_large_herm}   
Let $\lambda_{i}(A_{\beta})$ be the $i$th largest eigenvalue of $A_{\beta}$, for any fixed $1 \llq i \llq k$. Then, as $\beta \rightarrow \infty$,
\[
\lambda_i(A_{\beta}) \rightarrow \frac{1}{\sqrt{2k}} h_i^{(k)}~.
\]
Moreover, as $\beta \rightarrow \infty$,
\begin{eqnarray*}
\sqrt{\beta}\left(\lambda_1(A_{\beta}) -\frac{1}{\sqrt{2k}} h_1^{(k)} , \lambda_2(A_{\beta})-\frac{1}{\sqrt{2k}} h_2^{(k)}, \ldots, \lambda_k(A_{\beta})-\frac{1}{\sqrt{2k}} h_k^{(k)} \right) \rightarrow \frac{1}{\sqrt{2k}}~G 
\end{eqnarray*}
where $G \equiv (G_1, G_2, \ldots, G_k)$ is a $k$-variate Gaussian with covariance matrix 
\[
\mbox{Cov}(G_i,~G_j) ~=~ \frac{\sum\limits_{l=0}^{k-1} H_l^2(h_i^{(k)})H_l^2(h_j^{(k)}) + \sum\limits_{l=0}^{k-2}H_{l+1}(h_i^{(k)}) H_l(h_i^{(k)})H_{l+1}(h_j^{(k)}) H_l(h_j^{(k)})}{\left( \sum\limits_{l=0}^{k-1} H_l^2(h_i^{(k)}) \right)\left( \sum\limits_{l=0}^{k-1} H_l^2(h_j^{(k)}) \right) }~.
\]    
\end{theorem}

\begin{proof} Let $H$ be the $k \times k$ symmetric tridiagonal matrix
\begin{eqnarray} \label{fixed_h_matrix}
H = \frac{1}{\sqrt{2}} \left( \begin{array}{cccccc} 0 & \sqrt{k-1} & &&  & \\
				\sqrt{k-1} & 0 & \sqrt{k-2} &  & &\\
				           & \sqrt{k-2} & 0 &  & &\\
					   & & &\ddots & &\\
					   & & & &0 & \sqrt{1} \\
					   &  && &\sqrt{1} & 0 \end{array} \right )~.
\end{eqnarray}  

This matrix is the tridiagonal matrix corresponding to the $3$-term recurrence for Hermite polynomials. Its eigenvalues are the roots of the $k$th Hermite polynomial $H_k(x)$ (recall that we denoted them by $h_1^{(k)}, \ldots, h_k^{(k)}$), and the eigenvector corresponding to the $i$th eigenvalue $h_i^{(k)}$ is 
\[
v_i =  \left( \begin{array}{c} H_{k-1}(h_i^{(k)}) \\ H_{k-2}(h_i^{(k)}) \\ \vdots \\ H_1(h_i^{(k)}) \\ H_0(h_i^{(k)}) \end{array} \right)~.
\]


\begin{lemma} \label{fixed+gaussian_herm}
Let $A_{\beta}$ be as defined in the beginning of this section. Then almost surely 
\[
\lim_{\beta \rightarrow \infty} \sqrt{2 k \beta} A_{\beta} - \sqrt{\beta} H = Z~,
\]
where $Z$ is a tridiagonal matrix with standard normal variables on the diagonal and normal variables of mean $0$ and variance $1/4$ on the subdiagonal. All normal variables in $Z$ are mutually independent, subject only to the symmetry.
\end{lemma}

From now on we use the notation 
\begin{eqnarray} \label{def_z}
Z = \left ( \begin{array}{cccccc} M_k & N_{k-1} & & & & \\ N_{k-1} & M_{k-1} & N_{k-2} & & & \\ & N_{k-2} & M_{k-2} & & & \\  & & &\ddots & &\\  & & & & M_{2} & N_1\\ & & & & N_1 & M_1 \end{array} \right )~,
\end{eqnarray}
with $M_i$ being independent standard normals, while $N_i$ are independent normals of mean $0$ and variance $1/4$; all normal variables in $Z$ are mutually independent, subject only to the symmetry.

Lemma \ref{fixed+gaussian_herm} follows immediately from Lemma \ref{approx_chi}, since we are dealing with a finite number ($k-1$) of $\chi$ variables on the sub-diagonal of $A_{\beta}$, each converging in distributions to a normal variable.

Hence we have that, entry by entry, 
\[
A_{\beta} = \frac{1}{\sqrt{2k}} H + \frac{1}{\sqrt{2k \beta}} Z + o \left(\frac{1}{\sqrt{\beta}}\right)~,
\]
in distributions, as $\beta \rightarrow \infty$. 

Thus all zero- and first-order properties of $A_{\beta}$ are the same as for the random matrix $\frac{1}{\sqrt{2k}} H + \frac{1}{\sqrt{2k \beta}} Z$, where $Z$ as as above. In particular, for any $1\llq i \llq k$,
\[
\lambda_i(A_{\beta}) = \lambda_i \left(\frac{1}{\sqrt{2k}} H + \frac{1}{\sqrt{2k \beta}} Z \right) + o \left(\frac{1}{\sqrt{\beta}} \right)~,
\]
in distributions as $\beta \rightarrow \infty$.

Finally, with the help of perturbation theory Lemma \ref{perturb_lemma}, we obtain that for any $1\llq i \llq k$,
\[
\lambda_i(A_{\beta}) = \frac{1}{\sqrt{2k}} h_i^{(k)} + \frac{1}{\sqrt{2k \beta}} \frac{v_i^{T} Z v_i}{v_i^{T} v_i}\psi_i + o \left(\frac{1}{\sqrt{\beta}} \right)~,
\]
in distributions as $\beta \rightarrow \infty$.

Hence, using the notation \eqref{def_z}, 
\[
\sqrt{\beta} \left( \lambda_i(A_{\beta}) -  \frac{1}{\sqrt{2k}} h_i^{(k)} \right) = \frac{\sum_{l=0}^{k-1} H_l^2 (h_i^{(k)})~M_{l} ~+~ \sum_{l=1}^{k-1} H_l(h_i^{(k)}) H_{l-1}(h_i^{(k)})~N_l }{\sum_{l=0}^{k-1} H_l^2 (h_i^{(k)})}~ +~ o(1)~.
\]

The statement of Theorem \ref{beta_large_herm} follows. \end{proof}

Letting $k \rightarrow \infty$ in Theorem \ref{beta_large_herm}, we obtain the Corollary below.

\begin{corollary} \label{corr1}
Let $A_{\beta}$ be a matrix from the $k \times k$ $\beta$-Hermite ensemble, scaled by $\frac{1}{\sqrt{2k\beta}}$, and let $\lambda_1(A_\beta)$ be the largest eigenvalue of $A_{\beta}$. Then
\begin{eqnarray*} 
\lim_{n \rightarrow \infty} \lim_{\beta \rightarrow \infty} k^{-2/3} \left( \lambda_1(A_{\beta}) - 1\right ) \rightarrow \frac{a_1}{2} + \sigma^2 G,
\end{eqnarray*}
where $a_1 = -2.38810 \ldots$ is the largest root of the Airy {\tt Ai} function (see \cite{Abr-Steg}), and
\begin{eqnarray*}
\sigma^2 = 2\frac{\int_{0}^{\infty} {\tt Ai}^4(x+a_0)~dx}{\left(\int_{0}^{\infty} {\tt Ai}^2(x+a_0)~dx \right)^2}~ \sim ~ 0.41050 \ldots~~~.
\end{eqnarray*}
\end{corollary}

\begin{proof}
The corollary follows by using the special properties of the Hermite polynomial roots and the Airy function, as in \cite{Abr-Steg} and \cite[(1.81), (6.32)]{szego}, and from the fact that the matrix \eqref{fixed_h_matrix} is a discretization of the Airy operator, and as $k \rightarrow \infty$, the first eigenvector of the matrix becomes the {\tt Ai} solution of the operator which is bounded at $\infty$ and positive.
\end{proof}

\section{$ \beta$-Laguerre: zero and first-order approximations} \label{eigs_lag}

Let $k$ be fixed. Given a fixed $\gamma>0$, let $l_1^{(k)}, \ldots, l_k^{(k)}$ be the roots of the $k$th Laguerre polynomial of parameter $\gamma-1$, $L^{\gamma-1}_k$. Note that for any $\gamma \geq 0$, the Laguerre polynomials $L_{0}^{\gamma}, L_{1}^{\gamma}, \ldots$ are orthonormal with respect to the weight $x^{\gamma} e^{-x}$ on $[0, \infty)$; for $\gamma \in [-1,0)$ they admit formal definitions.

Let $B_{\beta}$ be a random matrix from the $\beta$-Laguerre ensemble of size $k$ and parameter $a_{\beta}$, scaled by $1/k\beta$. For the remainder of this section, we think of $\beta$ as a parameter. Suppose that, as $\beta$ grows large, 
\[
\lim_{\beta \rightarrow \infty} \frac{a_{\beta}}{\beta} = \frac{1}{2}(k+\gamma-1)~.
\]

Note that the requirement $a_{\beta} \!>\! (k-1) \beta/2$ constrains $\gamma$ to be positive.

\begin{theorem} \label{beta_large_lag}
Let $\lambda_{i}(B_{\beta})$ be the $i$th largest eigenvalue of $B_{\beta}$, for any fixed $1 \llq i \llq k$. Then, as $\beta \rightarrow \infty$,
\[
\lambda_i(B_{\beta}) \rightarrow \frac{1}{k} l_i^{(k)}~.
\]
Moreover, as $\beta \rightarrow \infty$,
\begin{eqnarray*}
\sqrt{\beta} \left( \lambda_1(B_{\beta}) - \frac{1}{k} l_1^{(k)}, ~\lambda_2(B_{\beta}) - \frac{1}{k} l_2^{(k)}, \ldots, \lambda_k(B_{\beta}) - \frac{1}{k} l_k^{(k)} \right) \rightarrow \frac{1}{k} G~,
\end{eqnarray*}
where $G \equiv (G_1, G_2, \ldots, G_k)$ is a centered $k$-variate Gaussian of covariance matrix 
\[
\mbox{Cov}(G_i,~G_j) ~=~ 2 \frac{(\gamma+k-1) (L^{\gamma}_{k-1}(l_i^{(k)}))^2 (L^{\gamma}_{k-1}(l_j^{(k)}))^2 + A_k(i,j) + B_k(i,j) +C_k(i,j)+D_k(i,j)}{\left( \sum \limits_{l=0}^{k-1} (L_{l}^{\gamma}(l_i^{(k)}))^2 \right) \left( \sum \limits_{l=0}^{k-1} (L_{l}^{\gamma}(l_j^{(k)}))^2 \right)}~
\]
\end{theorem}

\begin{eqnarray*}
A_k(i,j) & = & \sum\limits_{l=1}^{k-1} (\gamma+2(k-l)-1) (L^{\gamma}_{k-l-1} (l_i^{(k)}))^2   L^{\gamma}_{k-l-1} (l_j^{(k)}))^2 ~, \\
B_k(i,j) & = & \sum\limits_{l=1}^{k-1} (\gamma+2(k-l)) L^{\gamma}_{k-l-1}(l_i^{(k)}) L^{\gamma}_{k-l-1}(l_j^{(k)}) L^{\gamma}_{k-l}(l_i^{(k)})L^{\gamma}_{k-l}(l_j^{(k)})~,  \\
C_k(i,j) & = & \sum\limits_{l=1}^{k-1} \sqrt{\gamma\!+\!k\!-\!l\!}~\sqrt{k\!-\!l\!} ~~~\left( (L^{\gamma}_{k\!-l-\!1}(l_i^{(k)}))^2 L^{\gamma}_{k\!-l-\!1}(l_j^{(k)}) L^{\gamma}_{k\!-l}(l_j^{(k)}) ~+~ \right .\\
& & \left .~~~~~~~~~~~+~ (L^{\gamma}_{k\!-l-\!1}(l_j^{(k)}))^2 L^{\gamma}_{k\!-l-\!1}(l_i^{(k)}) L^{\gamma}_{k\!-l}(l_i^{(k)})\right)~, ~~\mbox{and} \\
D_k(i,j) & = & \sum\limits_{l=1}^{k-1} \sqrt{\gamma\!+\!k\!-\!l\!}~\sqrt{k\!-\!l\!} ~~~\left( (L^{\gamma}_{k\!-l}(l_i^{(k)}))^2 L^{\gamma}_{k\!-l-\!1}(l_j^{(k)}) L^{\gamma}_{k\!-l}(l_j^{(k)}) ~+~ \right .\\
& & \left .~~~~~~~~~~~+~ (L^{\gamma}_{k\!-l}(l_j^{(k)}))^2 L^{\gamma}_{k\!-l-\!1}(l_i^{(k)}) L^{\gamma}_{k\!-l}(l_i^{(k)})\right)~.
\end{eqnarray*}

\begin{proof} The proof follows in the footsteps of that of Theorem \ref{beta_large_herm}.

Let $L_{\gamma}$ be the $k \times k$ (symmetric) positive definite matrix
\footnotesize{
\begin{eqnarray} \label{fixed_l_matrix}
L_{\gamma} = \left ( \begin{array}{ccccc} \gamma+k-1 & \sqrt{\gamma+k-1} \sqrt{k-1} & & &\\
					  \sqrt{\gamma+k-1} \sqrt{k-1} & 2(k-2)+\gamma+1   & \sqrt{\gamma+k-2} \sqrt{k-2} & &\\
& \sqrt{\gamma+k-2} \sqrt{k-2} &  2(k-3)+\gamma+1 & & \\
& & \ddots & & \\
& & & \sqrt{\gamma+2} \sqrt{2} &\\
& & \sqrt{\gamma+2} \sqrt{2} & 3+\gamma & \sqrt{\gamma+1} \sqrt{1} \\
& & & \sqrt{\gamma+1} \sqrt{1} & 1+\gamma  
\end{array} \right )
\end{eqnarray}}

\normalsize 
We can write $L_{\gamma} = B_{\gamma} B_{\gamma}^{T}$, with
\footnotesize{
\begin{eqnarray}
B_{\gamma} = \left( \begin{array}{ccccc} \sqrt{\gamma+k-1} & & & &\\
					\sqrt{k-1} & \sqrt{\gamma-k} & & & \\
						   & \ddots & \ddots & & \\
						   &   & \sqrt{2} & \sqrt{\gamma+1} & \\
						   & & & \sqrt{1} & \sqrt{\gamma} \end{array} \right)~.
\end{eqnarray}}

\normalsize
The matrix $L_{\gamma}$ has as eigenvalues the roots of the $k$th Laguerre polynomial of parameter $\gamma-1$, $L_k^{\gamma-1}(x)$ (recall that we have denoted them by $l_1^{(k)}, \ldots, l_k^{(k)}$), and an eigenvector corresponding to the $i$th eigenvalue $l_i^{(k)}$ is 
\[
w_i = \left ( \begin{array}{c} L_{k-1}^{\gamma} (l_i^{(k)}) \\ L_{k-2}^{\gamma} (l_i^{(k)}) \\ \vdots \\ L_{1}^{\gamma} (l_i^{(k)}) \\ L_{0}^{\gamma} (l_i^{(k)}) \end{array} \right)~.
\]

We define $\phi_i \equiv w_i / ||w_i||_2$ to be a length $1$ eigenvector corresponding to the $i$th eigenvalue $l_i$.

\begin{lemma} \label{fixed+Gaussian_term}
Let $B_{\beta}$ be as in the statement of Theorem \ref{beta_large_lag}. Then 
\[
\lim_{\beta \rightarrow \infty} k \beta B_{\beta} - \beta L^{\gamma} = \frac{1}{\sqrt{2}} (B_{\gamma} Z^{T} + ZB_{\gamma}^{T}) ~,
\]
in distributions, where $Z$ is a lower bidiagonal matrix with standard normal variables on the diagonal and on the subdiagonal. All normal variables in $Z$ are mutually independent, subject only to the symmetry constraint.
\end{lemma}

We use the notation
\footnotesize{
\begin{eqnarray} \label{def_newz}
Z \equiv \left ( \begin{array}{ccccc} M_k & & & & \\
				      N_{k-1} & M_{k-1} & & & \\
				 	& \ddots & \ddots & & \\
					&   &	N_2 & M_2 & \\
					& & & N_1 & M_1 \end{array} \right )~.
\end{eqnarray}}

\normalsize

Once again, the proof for this lemma follows from the construction of the Laguerre matrix as a lower bidiagonal random matrix times its transpose, and from Lemma \ref{approx_chi} applied to the $\chi$ entries on the bidiagonal random matrix (there is a finite number $2k-1$ of them).

Just as in the Hermite case, Lemma \ref{fixed+Gaussian_term} allows us to write that, entry by entry,
\[
B_{\beta} = \frac{1}{k} L_{\gamma} + \frac{1}{k \sqrt{2\beta}} (B_{\gamma} Z^{T} + ZB_{\gamma}^{T}) + o \left( \frac{1}{\sqrt{\beta}} \right)~,
\] in distributions, as $\beta \rightarrow \infty$. Thus once again, 
\[
\lambda_i(B_{\beta}) = \lambda_i( \frac{1}{k} L_{\gamma} + \frac{1}{k\sqrt{2\beta}} (B_{\gamma} Z^{T} + ZB_{\gamma}^{T})) + o \left( \frac{1}{\sqrt{\beta}} \right)~,
\] in distributions, as $\beta \rightarrow \infty$.

Finally, perturbation theory Lemma \ref{perturb_lemma} applies once again to yield that, as $\beta \rightarrow \infty$, 
\[
\lambda_i(B_{\beta}) = \frac{1}{k} l_i^{(k)} + \frac{1}{k \sqrt{2\beta}} \frac{w_i^{T} (B_{\gamma} Z^{T} + ZB_{\gamma}^{T}))w_i}{w_i^{T} w_i} + o \left( \frac{1}{\sqrt{\beta}} \right)
\] in distributions.

Since $w_i^{T} B_{\gamma} Z^{T} w_i = w_i^{T} Z B_{\gamma}^{T} w_i$, we can write
that, as $\beta \rightarrow \infty$, 
\[
\lambda_i(B_{\beta}) = \frac{1}{k} l_i^{(k)} + \frac{\sqrt{2}}{k \sqrt{\beta}} \frac{w_i^{T} B_{\gamma} Z^{T}w_i}{w_i^{T}w_i} + o \left( \frac{1}{\sqrt{\beta}} \right)
\] in distributions.

Thus, using notation \eqref{def_newz}, we can write
\begin{eqnarray*}
\sqrt{\beta} \left( \lambda_i(B_{\beta}) - \frac{1}{k} l_i^{(k)} \right) & = & \frac{\sqrt{2}}{k} \frac{ \sqrt{\gamma} (L_0^{\gamma}(l_i^{(k)}))^2 + \mbox{Sum}_1 + \mbox{Sum}_2}{\sum\limits_{l=0}^{k-1} L_l^{\gamma}(l_i^{(k)})^2}~,
\end{eqnarray*}
with
\begin{eqnarray*}
\mbox{Sum}_1 & = & \sum_{l=1}^{k-1} \left ( \sqrt{\gamma+l} (L_{l}^{\gamma}(l_i^{(k)}))^2 + \sqrt{l} L_{l}^{\gamma}(l_i^{(k)}) L_{l-1}^{\gamma}(l_i^{(k)}) \right ) M_{l+1} ~, ~~~\mbox{and} \\
\mbox{Sum}_2 & = & \sum_{l=1}^{k-1} \left ( \sqrt{\gamma+l} L_{l}^{\gamma}(l_i^{(k)}) L_{l-1}^{\gamma}(l_i^{(k)}) + \sqrt{l} (L_{l-1}^{\gamma}(l_i^{(k)}))^2 \right ) N_l~.
\end{eqnarray*}

The statement of the theorem follows.

\end{proof}

\section{Applications: Level densities} \label{applications}

We can compare the large $\beta$ asymptotics to the theoretical answer for the distribution of a randomly chosen eigenvalue. For large $n$, this is the well-know semicircle law (for the Hermite ensembles) or equivalent thereof (for Laguerre ensembles), but we are interested in finite $n$.

We found that even for $\beta$ small, the approximation can be quite reasonable. 

We summarize the large $\beta$ answer as a sum of Gaussians in Corollaries \ref{c1} and \ref{c2}.

\begin{corollary} \label{c1}
Let $k$ be fixed, and $f_{k, \beta}$ be the level density of the scaled (by $1/\sqrt{2k\beta}$) $k \times k$ $\beta$-Hermite\index{$\beta$-Hermite ensemble!asymptotics for $\beta$ large} ensemble. Let $g_{k, \beta}$ be as below:
\[
g_{k, \beta}(x) = \frac{1}{k} \sum_{i=1}^k \frac{1}{\sqrt{2 \pi} \sigma_i} e^{-\frac{(x - \mu_i)^2}{2\sigma_i^2}}~,
\]
where $\mu_i = \frac{h_i^{(k)}}{\sqrt{2k}}$ and $\sigma_i = \frac{1}{\sqrt{2k\beta}} \sqrt{\mbox{Var}(G_i)}$, with $h_i$ and Var$(G_i)$ as in Section \ref{eigs_herm}. Then for any $x$, 
\[
\lim_{\beta \rightarrow \infty} \sqrt{\beta}~(f_{k, \beta}(x) - g_{k, \beta}(x)) ~ = ~ 0~.
\]
\end{corollary}

\begin{corollary} \label{c2}
Let $k$ and $\gamma>0$ be fixed, and $f_{k, \beta, \gamma}$ be the level densit\index{level density!asymptotic, large $\beta$}y of the scaled (by $1/(k \beta)$) $k \times k$ $\beta$-Laguerre\index{$\beta$-Laguerre ensemble!asymptotics for $\beta$ large} ensemble of parameter $a = \frac{\beta}{2}(k-1+\gamma)$. Let $g_{k, \beta, \gamma}$ be as below:
\[
g_{k, \beta, \gamma}(x) = \frac{1}{k} \sum_{i=1}^k \frac{1}{\sqrt{2 \pi}\sigma_i} e^{-\frac{(x - \mu_i)^2}{2\sigma_i^2}}~,
\]
where $\mu_i = \frac{l_i^{(k)}}{k}$ and $\sigma_i = \frac{1}{k\sqrt{\beta}} \sqrt{\mbox{Var}(G_i)}$, with $l_i^{(k)}$ and Var$(G_i)$ as in Section \ref{eigs_lag}. Then for any $x$,
\[
\lim_{\beta \rightarrow \infty} \sqrt{\beta}~(f_{k, \beta, \gamma}(x) - g_{k, \beta, \gamma}(x)) ~ = ~ 0~.
\]
\end{corollary}

While these approximations are simple enough (a sum of Gaussians is an easily recognizable shape that is also easy to work with), one may wonder how big $\beta$ has to be in order for these approximations to become ``accurate'' (for example, in order to \emph{appear} accurate in a plot, the approximations have to be accurate to about 2-3 digits). We have found that, in either of the two cases, the answer is surprisingly low.

In the following two subsections, we have used only even integer values of $\beta$ for our plots, because (in addition to $\beta = 1$) those are the only ones for which (to the best of our knowledge) there are exact formulas for the level densities. The plots were obtained with the help of our Maple Library, \textbf{MOPs} (\textit{Multivariate Orthogonal Polynomials (symbolically)}), which was used for computing the orthogonal and Jack polynomial quantities involved; these were translated into polynomials which were then plotted in \textbf{MATLAB}. For a reference on MOPs see \cite{dumitriu04a}.

\subsection{Level density plots: the Hermite case}

In the following, we illustrate the accuracy of the level density approximation by a sum of Gaussians for $\beta$ relatively small ($4$ to $10$) by plotting them against the true level densities.

\begin{figure}[ht!]
\parbox[b]{5cm}{\epsfig{figure = 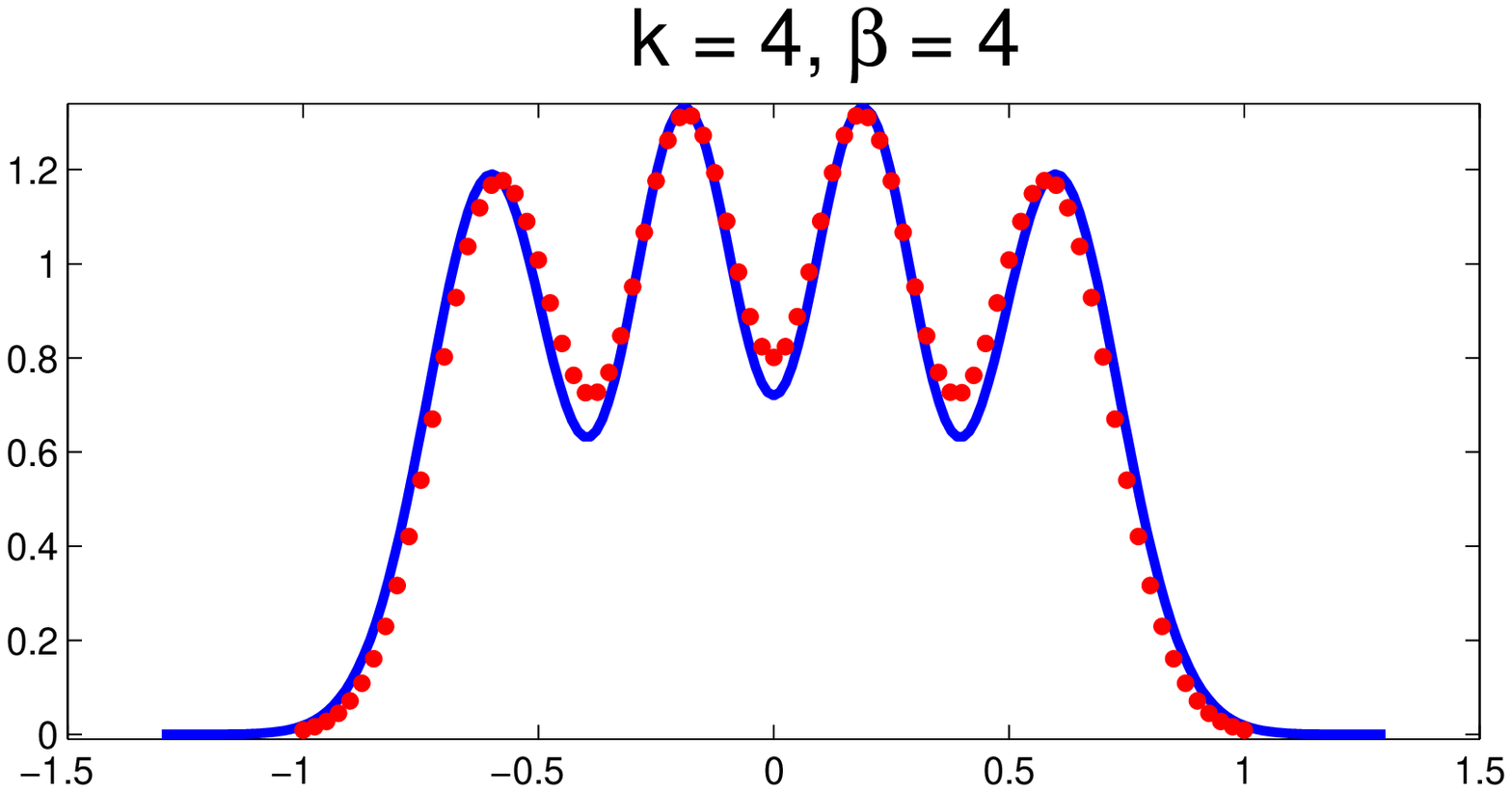, height = 3.75cm}} \hspace{2.5cm}
\parbox[b]{5cm}{\epsfig{figure = 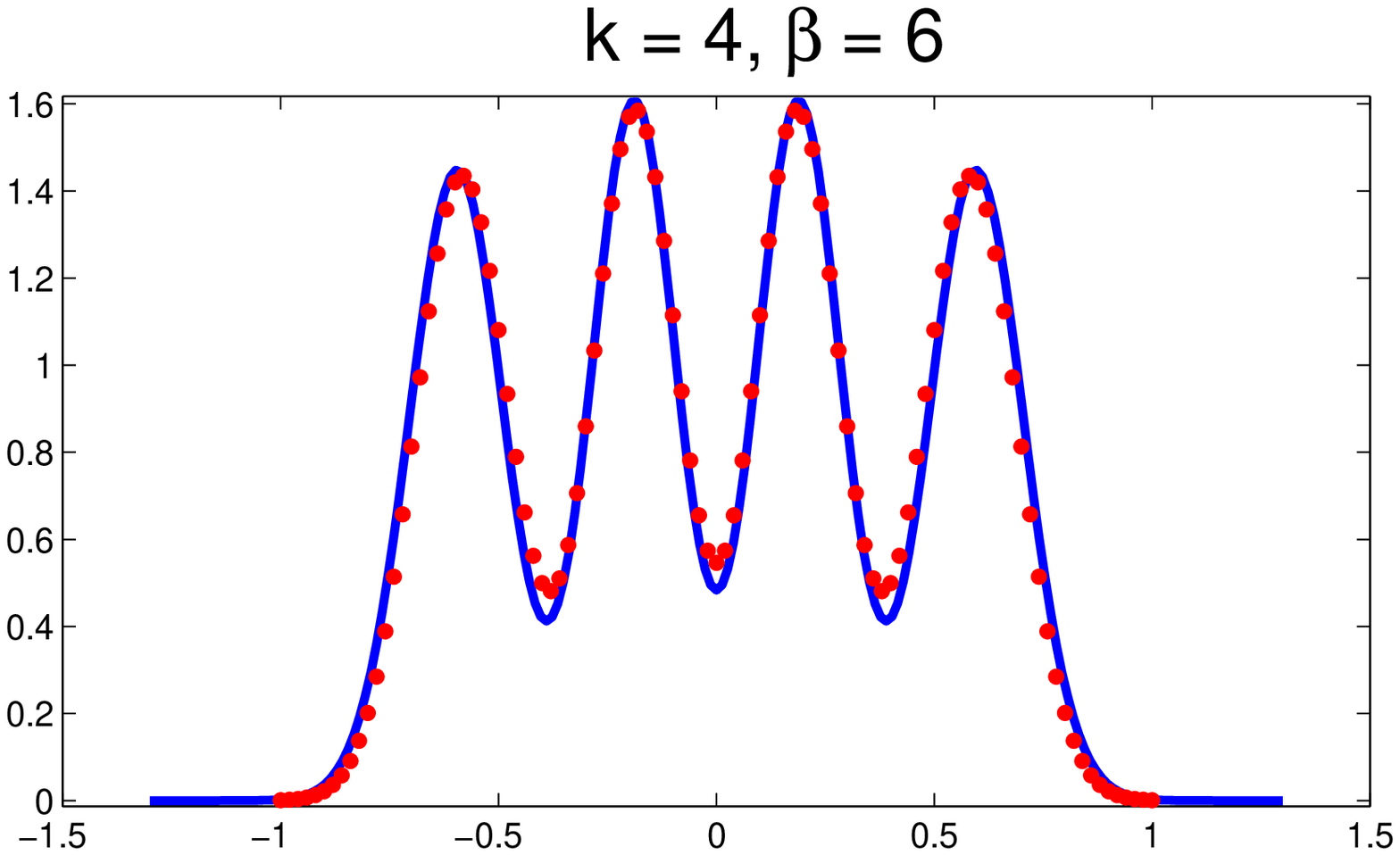, height = 4.4cm}} \\
\parbox[b]{6cm}{\epsfig{figure = 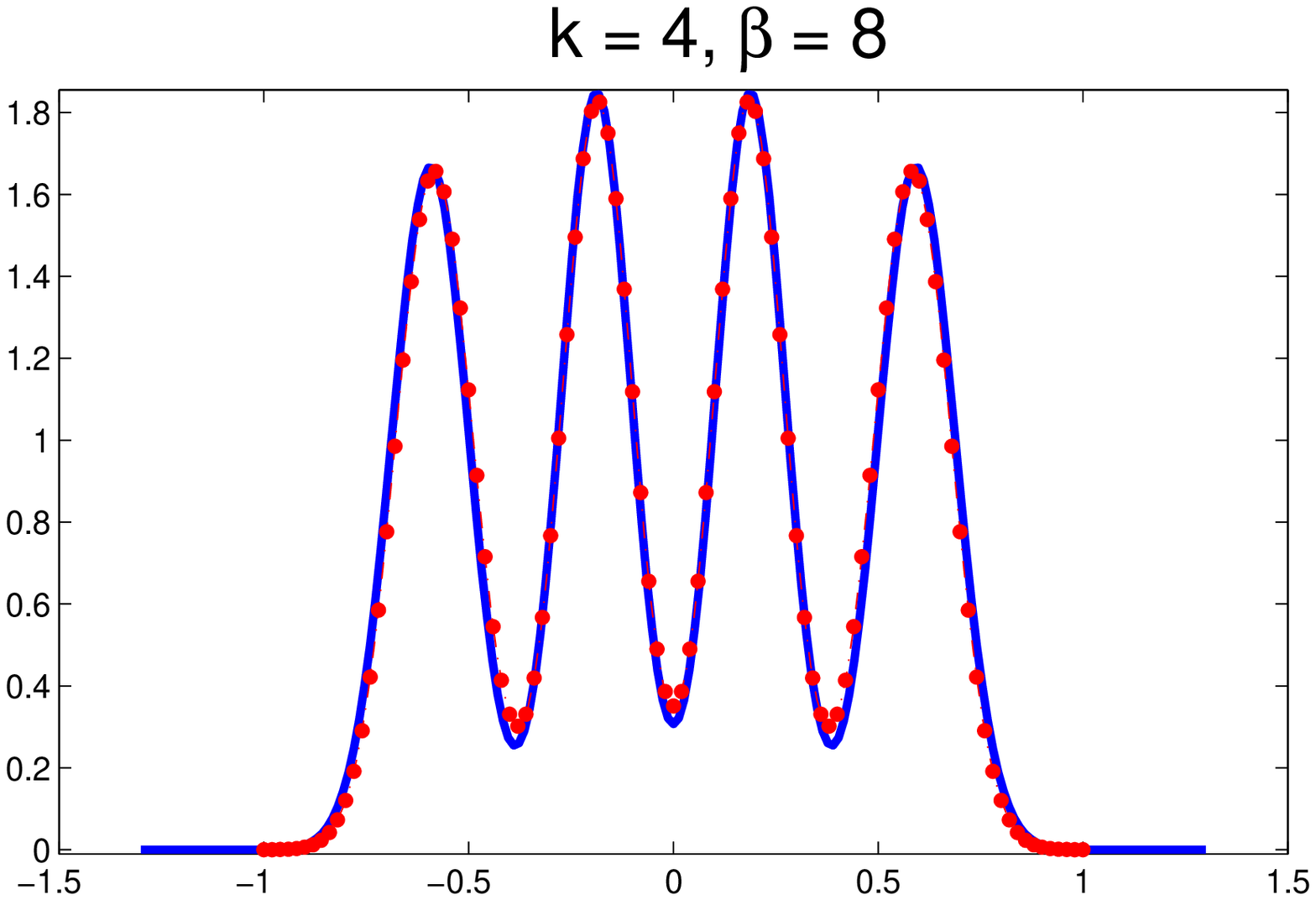, height = 4.9cm}} \hspace{1.5cm}
\parbox[b]{6cm}{\epsfig{figure = 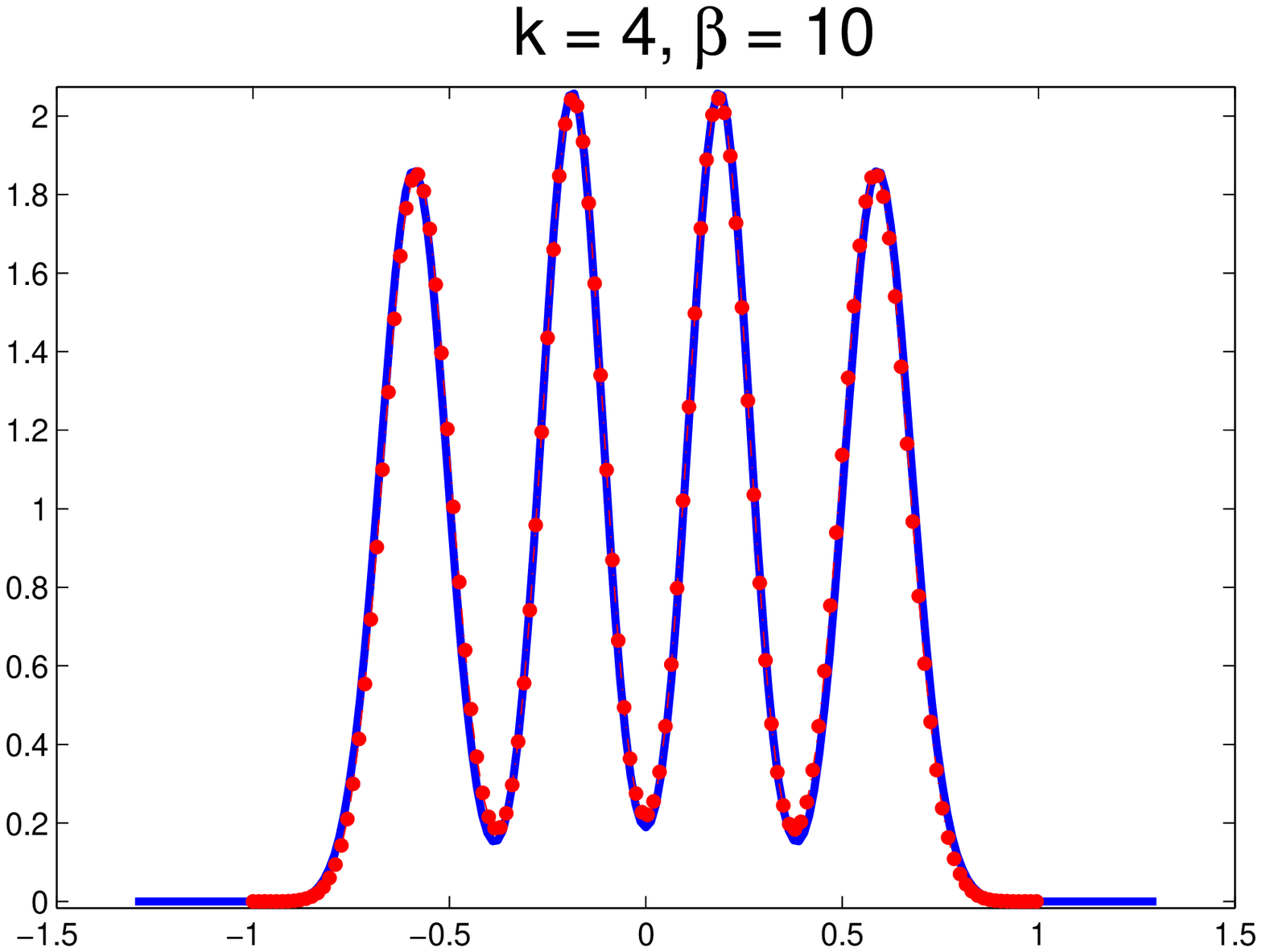, height = 5.4cm}}
\caption{Hermite case:  sum of Gaussians approximation to the level densities (dots) and exact level densities (lines) for $k = 4$, and $\beta = 4,6,8,10$}\label{lev4dens_4_10_herm}
\end{figure}

Figure \ref{lev4dens_4_10_herm} plots the Hermite case with $k=4$. 

In Figure  \ref{lev4dens_4_10_herm}, we let $k=4$, and gradually increase $\beta$ (from $4$ to $10$) to show how the plots become more and more similar. For $\beta = 10$, the two plots appear to overlap.

\begin{figure}[!ht]
\parbox[b]{5cm}{\epsfig{figure = 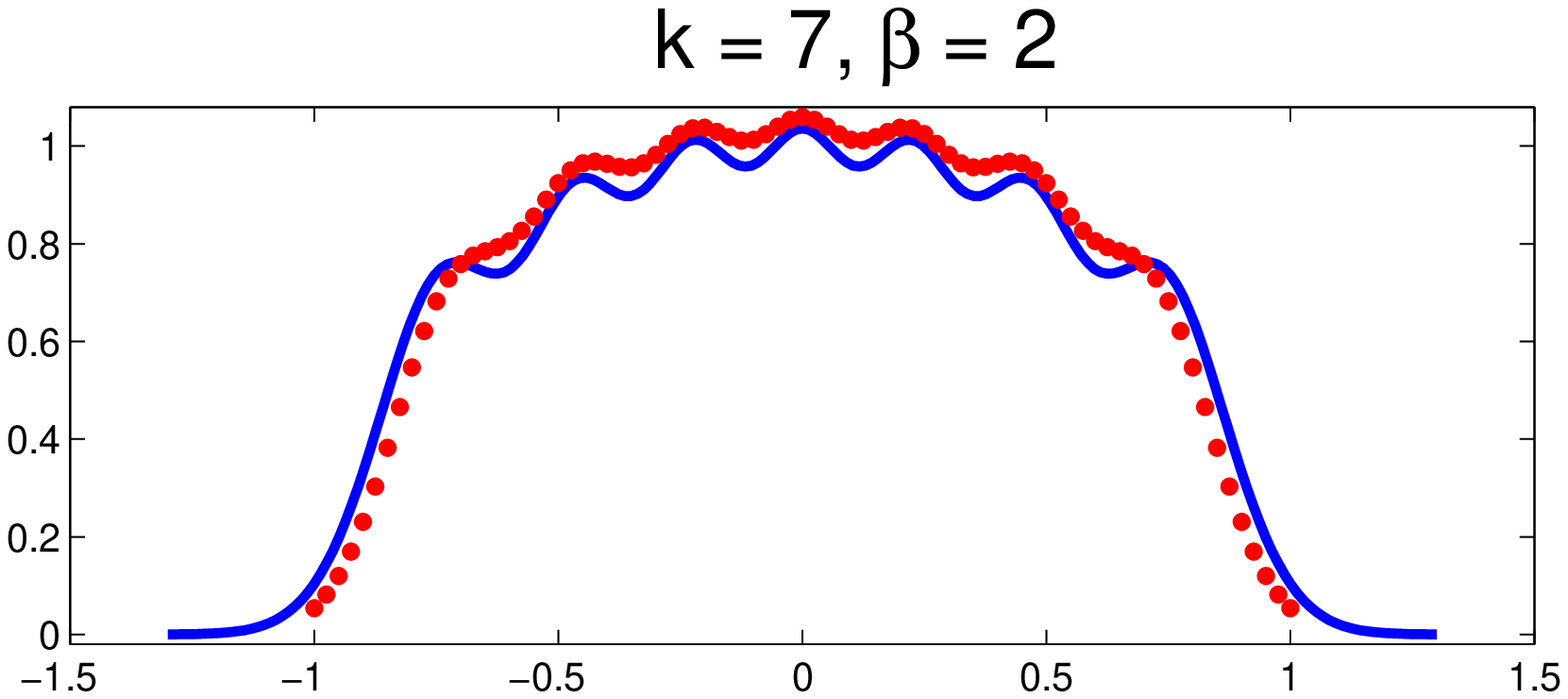, height = 3.2cm}} \hspace{2.4cm}
\parbox[b]{5cm}{\epsfig{figure = 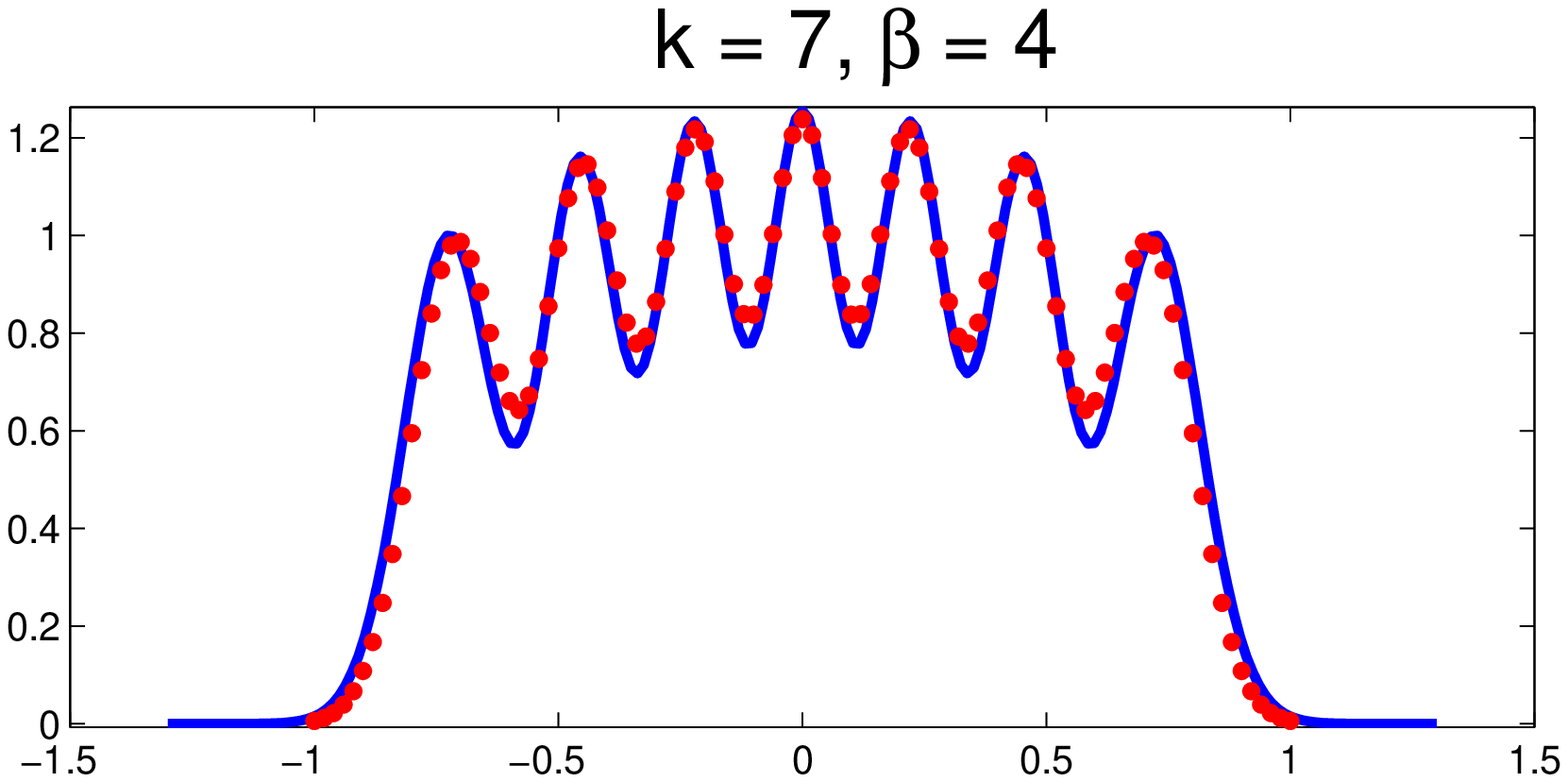, height = 3.67cm}} \\
\parbox[b]{5cm}{\epsfig{figure = 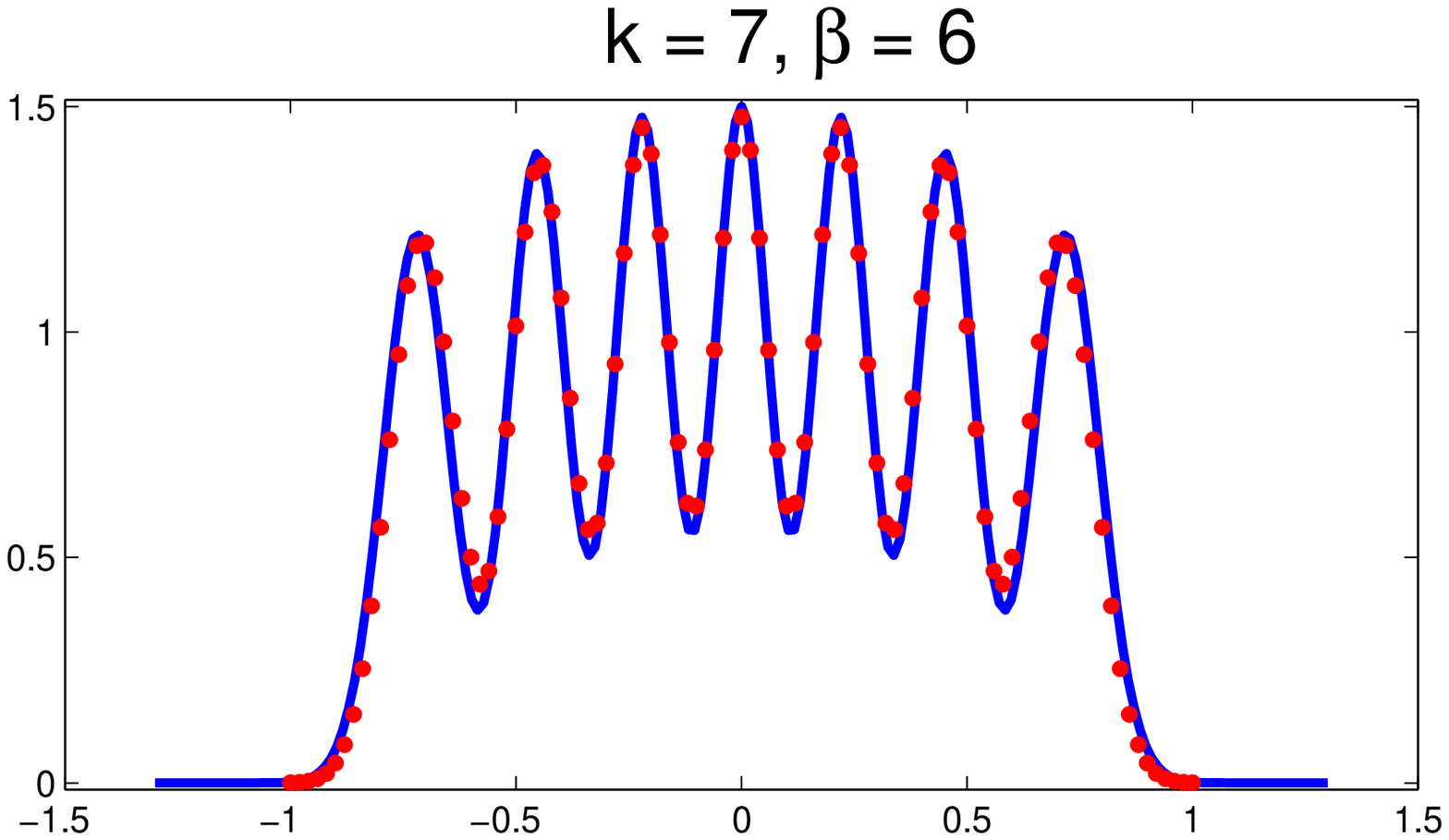, height = 4.2cm}} \hspace{2.6cm}
\parbox[b]{5cm}{}
\caption{Hermite case: sum of Gaussians approximation to the level densities (dots) and exact level densities (lines) for $k = 7$, and $\beta = 2,4,6$.} \label{lev7dens_2_6_herm}
\end{figure}

If we plot the densities for $k = 7$ (as in Figure \ref{lev7dens_2_6_herm}), $\beta = 6$ already provides a very good approximation. 

We can conclude that the approximation works well for low values of  $\beta$, in the Hermite case.

\subsection{Level densities: the Laguerre case}

In the Laguerre case, we cut the parameter cube with two different slices, as explained below. For plotting purposes we have considered $k=4$ in both.

To specify a Laguerre density, there is always an intrinsic parameter: the power ``\emph{p}'' (as in $x^{p} e^{-x}$). However, in this story, there are two Laguerre densities: one for the eigenvalue
p.d.f.\, that is in the Laguerre ensemble density, and a second (different!) one for the Laguerre polynomial corresponding to the limiting level density. We found this surprising at first.
In order to study $\beta \rightarrow \infty$ limits we looked at both possibilities
of holding a parameter constant, as depicted in the table below.

\vspace{.5cm}

\begin{center}
\begin{tabular}{c|c|c|c|c|c} 
& Fixed &  & Other & Eigenvalue & Limiting \\
& quantities & Variable & quantities & p.d.f.\ & Laguerre  \\
&&&&& polynomial\\ \hline &&&&&\\
\textit{\textbf{a)}} & $k$, {\bf $\gamma$} & $\beta \rightarrow \infty$ & $a = \frac{\beta}{2}(k+\gamma-1)$ & $c |\Delta|^{\beta} \prod_{i=1}^k \lambda_i^{\frac{\beta}{2}\gamma-1} e^{-\lambda_i/2}$ & $L^{\gamma-1}_k(x)$ \\ 
       &                        &                            & $p = \frac{\beta}{2}\gamma -1$ &                                                          &            \\ & & & & & \\ \hline &&&&&\\
\textit{\textbf{b)}} & $k$, {\bf $p$} & $\beta \rightarrow \infty$ & $a = p + \frac{\beta}{2}(k-1)$ & $c |\Delta|^{\beta} \prod_{i=1}^k \lambda_i^{p} e^{-\lambda_i/2}$ & $L^{-1}_k(x)$ \\
        &                       &                            & $\gamma = \frac{2}{\beta}(p+1)$                                                       & \\ 
\end{tabular}
\end{center}

\vspace{.5cm} 

\noindent \textit{\textbf{Case a)}.} This case holds $\gamma$ (and therefore the limiting Laguerre polynomial, whose roots are the limits of the scaled eigenvalues) constant as $\beta \rightarrow \infty$. 

Note that both the Laguerre ensemble parameter $a= \frac{\beta}{2}(k+\gamma-1)$ and the power $p= \gamma \frac{\beta}{2}-1 $ are increasing functions of $\beta$. 

By prescribing $\gamma$, in the limit as $\beta \rightarrow \infty$ the plot should become a sum of delta functions at the roots of the Laguerre polynomial $L^{\gamma-1}_k(x)$.

\begin{figure}[ht!]
\parbox[b]{6cm}{\epsfig{figure =  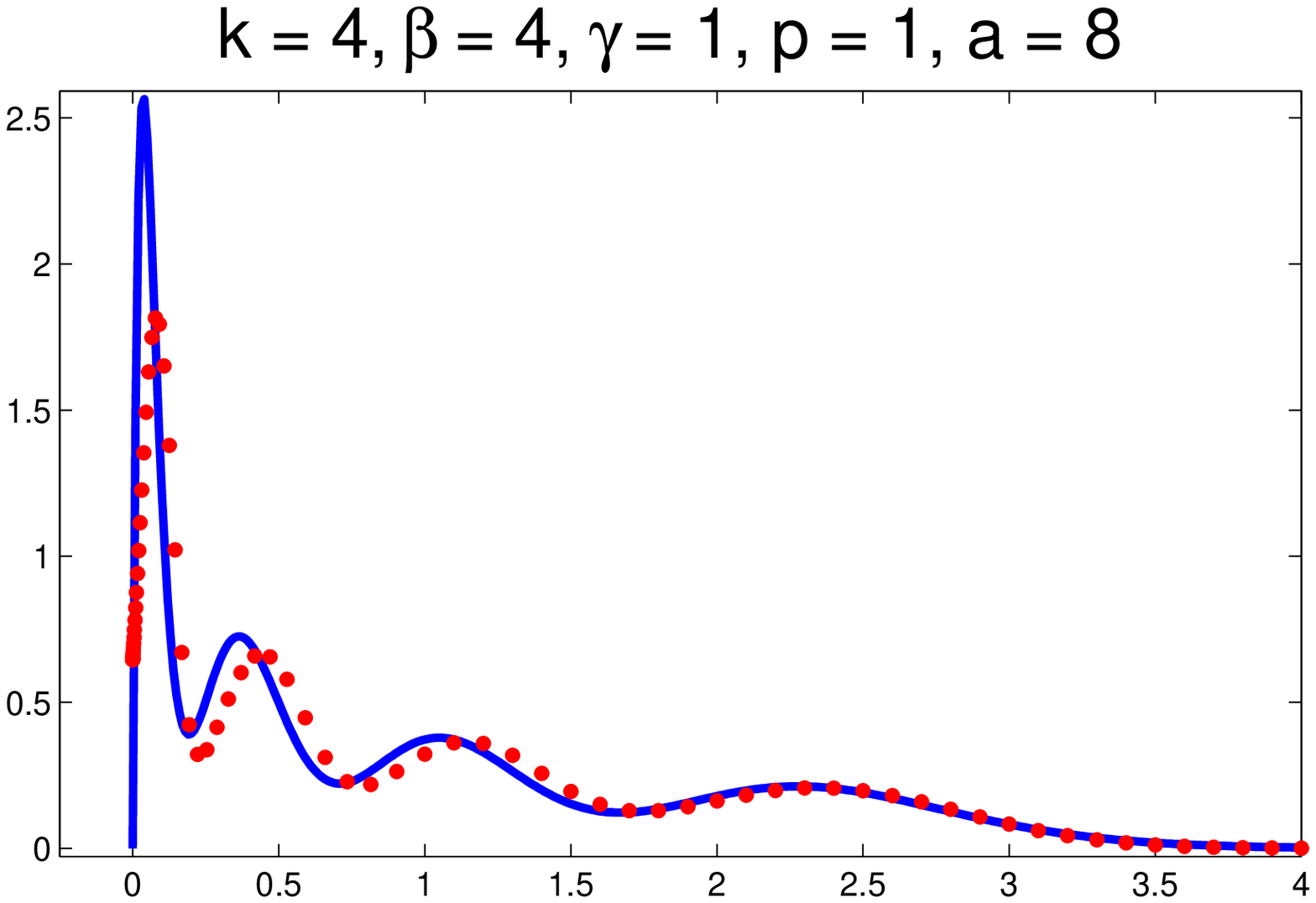, height = 4.75cm}} \hspace{1cm}
\parbox[b]{6cm}{\epsfig{figure =  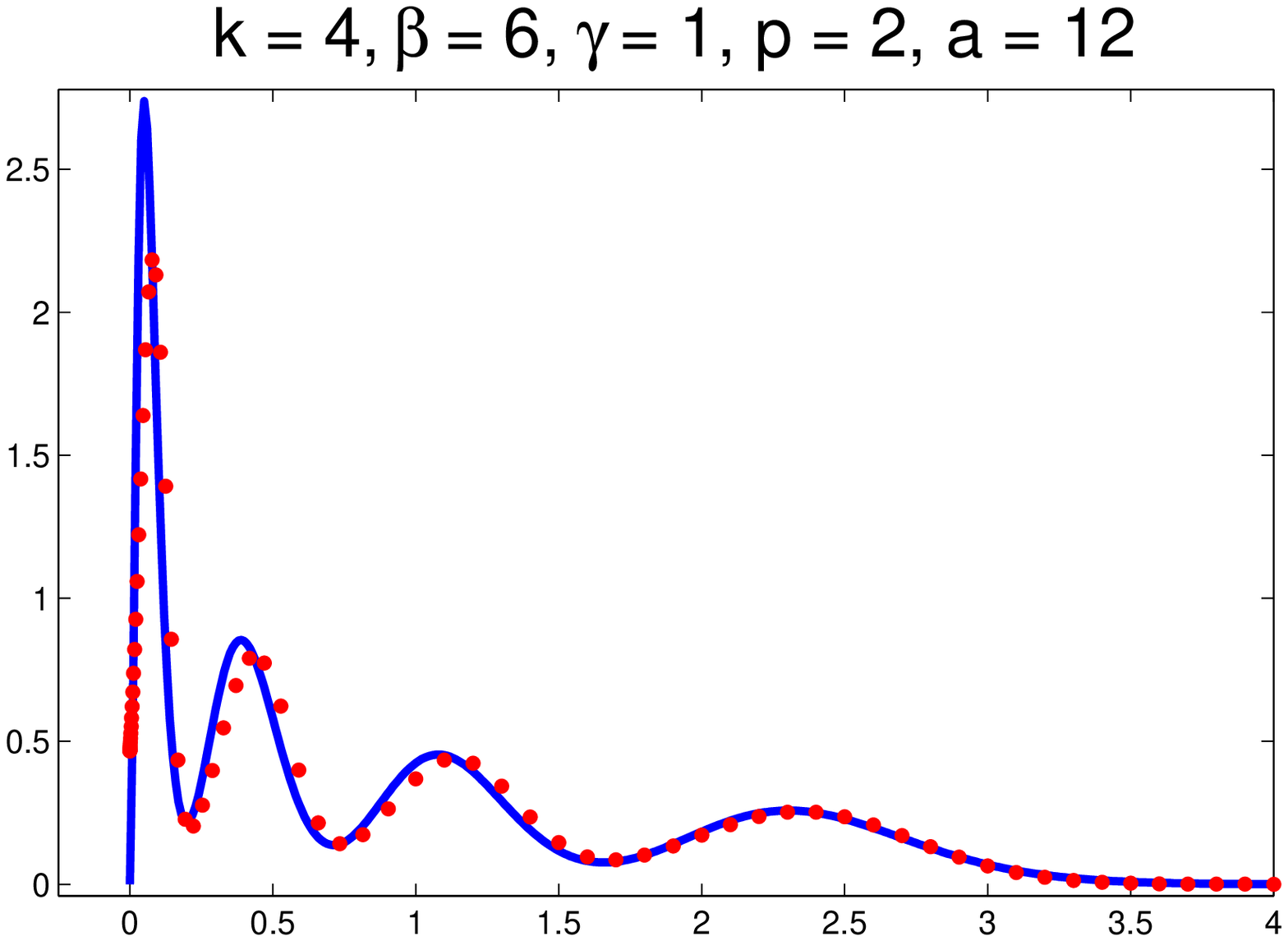, height = 5.11cm}} \\
\parbox[b]{6cm}{\epsfig{figure = 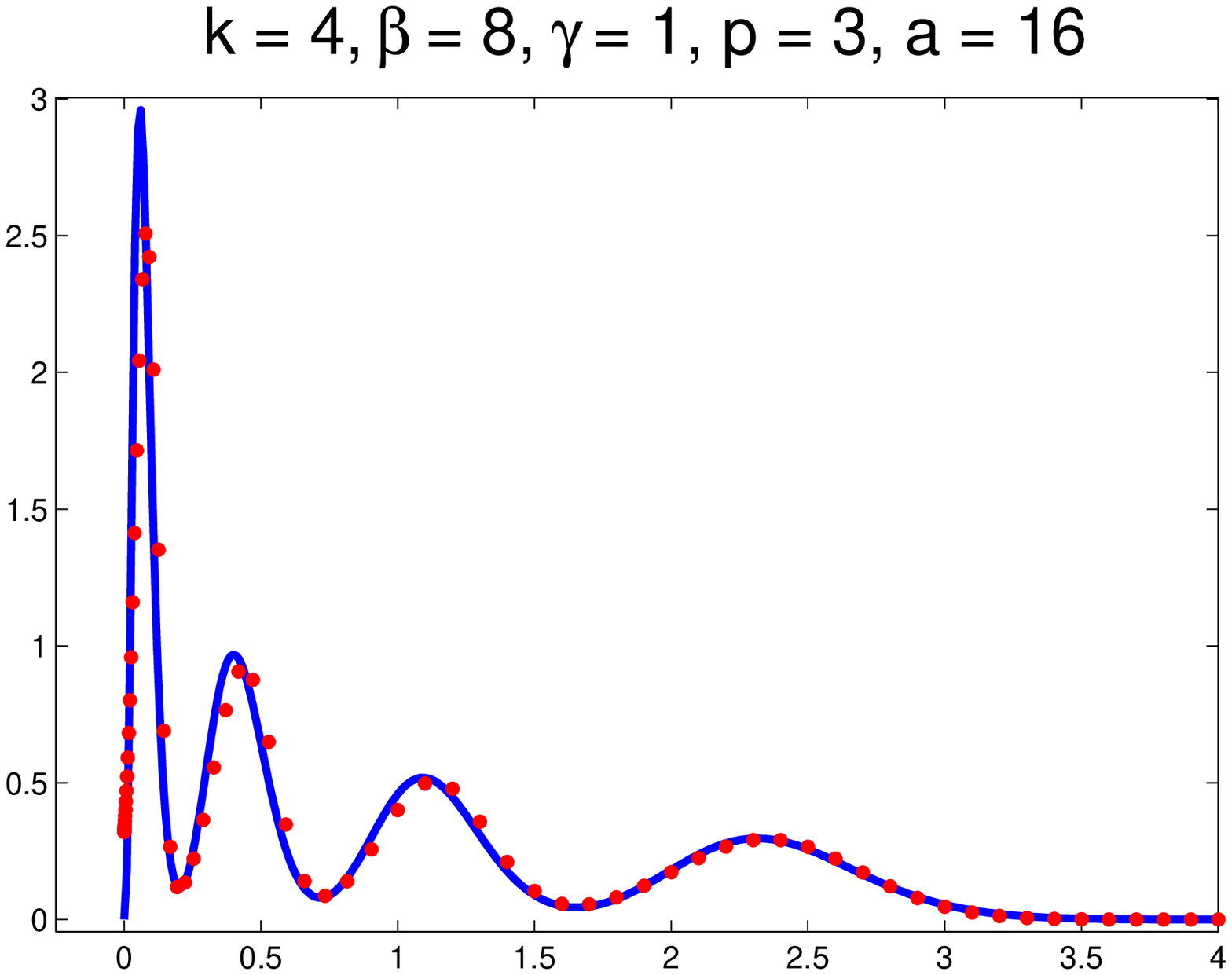, height = 5.45cm}} \hspace{1cm}
\parbox[b]{6cm}{\epsfig{figure = 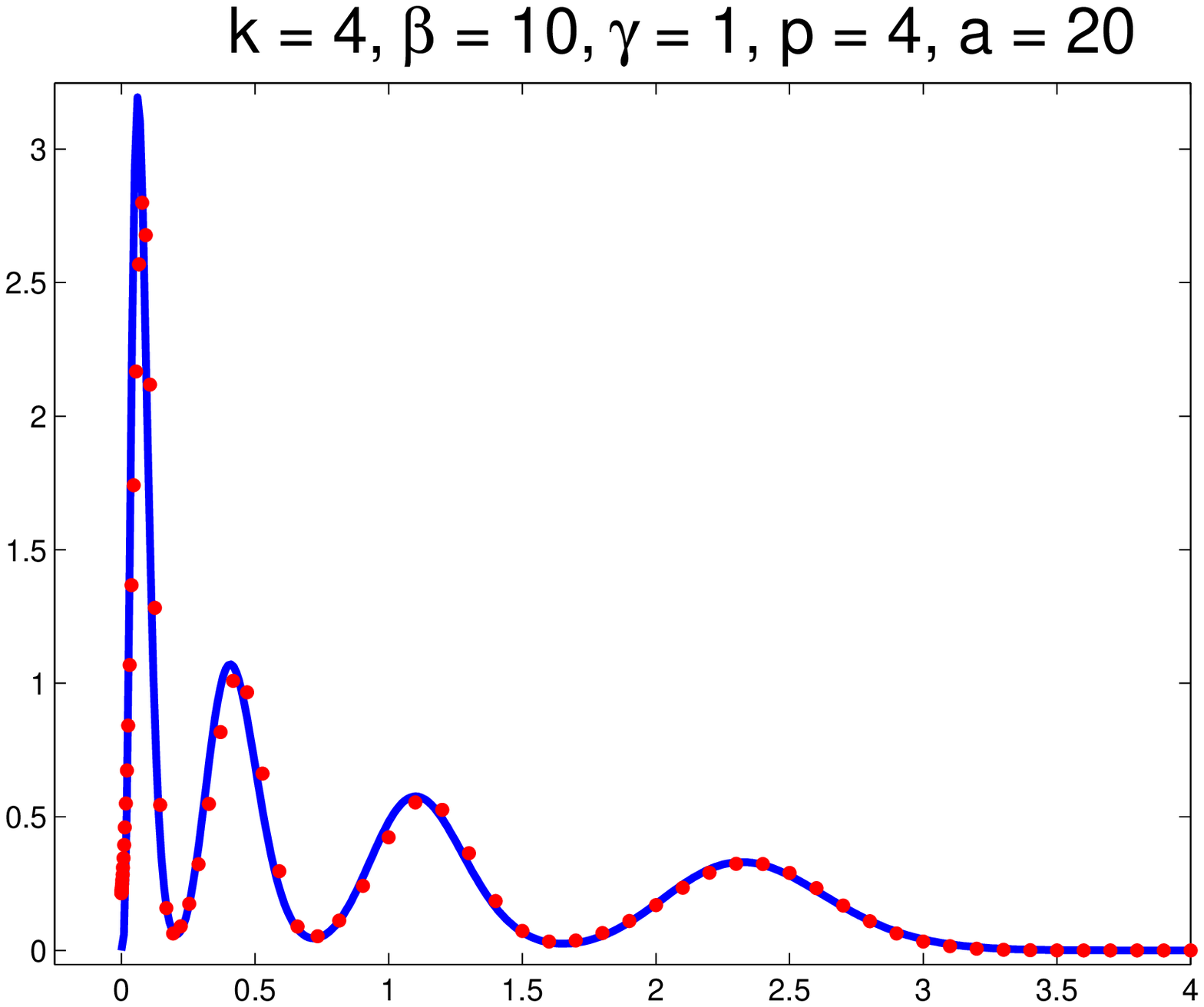, height = 5.84cm}}
\caption{Laguerre case a):  sum of Gaussians approximation to the level densities (dots) and exact level densities (lines) for $k = 4$, $\gamma = 1$, and $\beta = 4,6,8,10$}\label{levdenslag_g1}
\end{figure}

In Figure \ref{levdenslag_g1} we take $k = 4$, $\gamma = 1$, $\beta = 4, 6,8,10$, and $a = 8, 12, 16, 20$ (equivalently, $p = 1,2,3,4$). Note that the approximation is very good for $\beta = 10$.

\vspace{.5cm}

\noindent \textit{\textbf{Case b)}.} This case holds the power $p$ constant in the weight $|\Delta(\Lambda)|^{\beta} \prod_{i=1}^k \lambda_i^{p} e^{-\lambda_i/2}$, thereby changing the parameter $\gamma$ and the Laguerre polynomial. In this second test, as $\beta \rightarrow \infty$, $\gamma = \frac{2}{\beta}(p+1) ~\rightarrow 0$.

Thus as $\beta \rightarrow \infty$, the plot should become a sum of delta functions at the roots\index{Laguerre polynomial!roots} of the polynomial $L^{-1}_n(x)$.

\begin{figure}[ht!]
\parbox[b]{7.1cm}{\epsfig{figure =  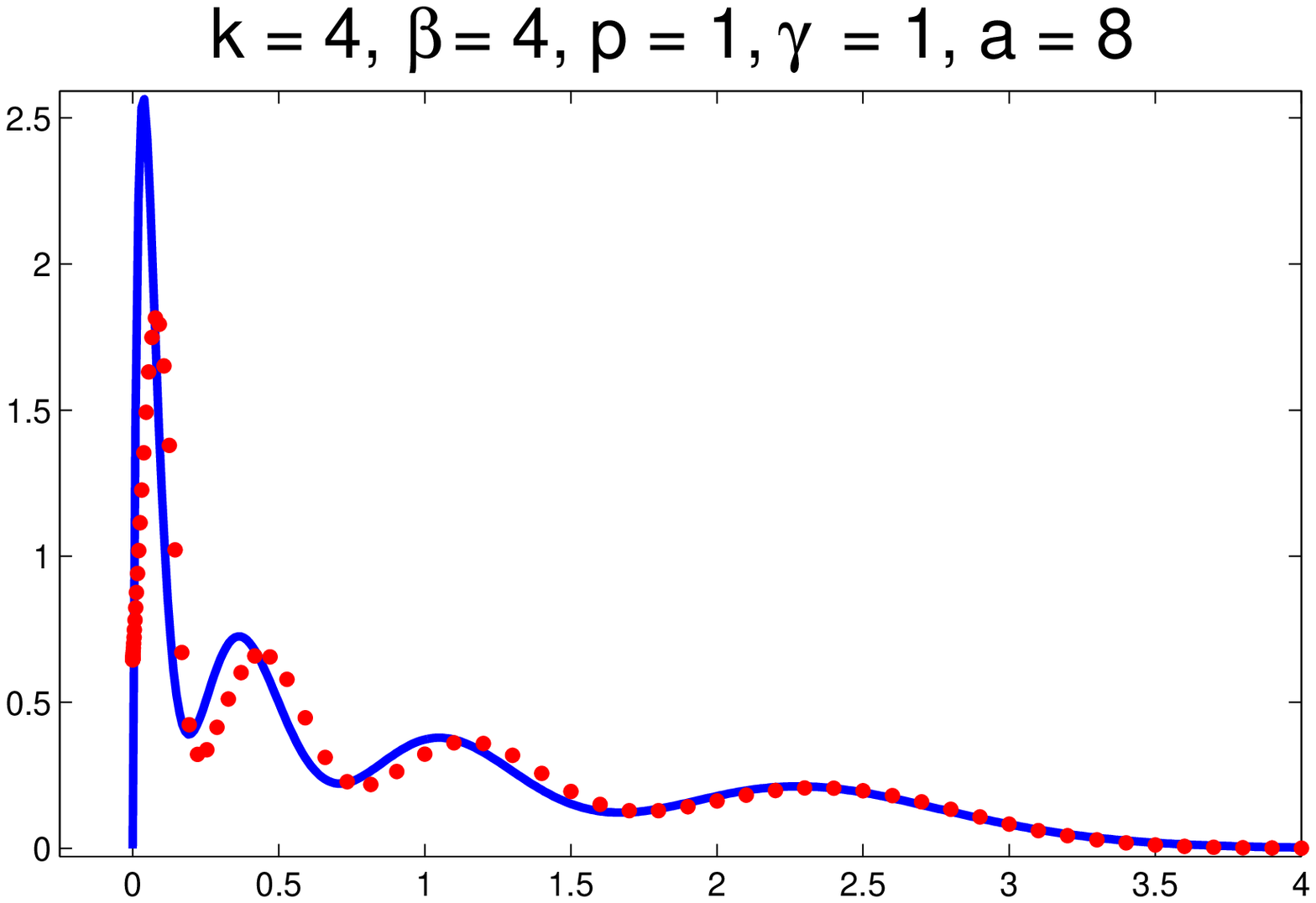, height = 4.5cm}} \hspace{.5cm}
\parbox[b]{7.1cm}{\epsfig{figure =  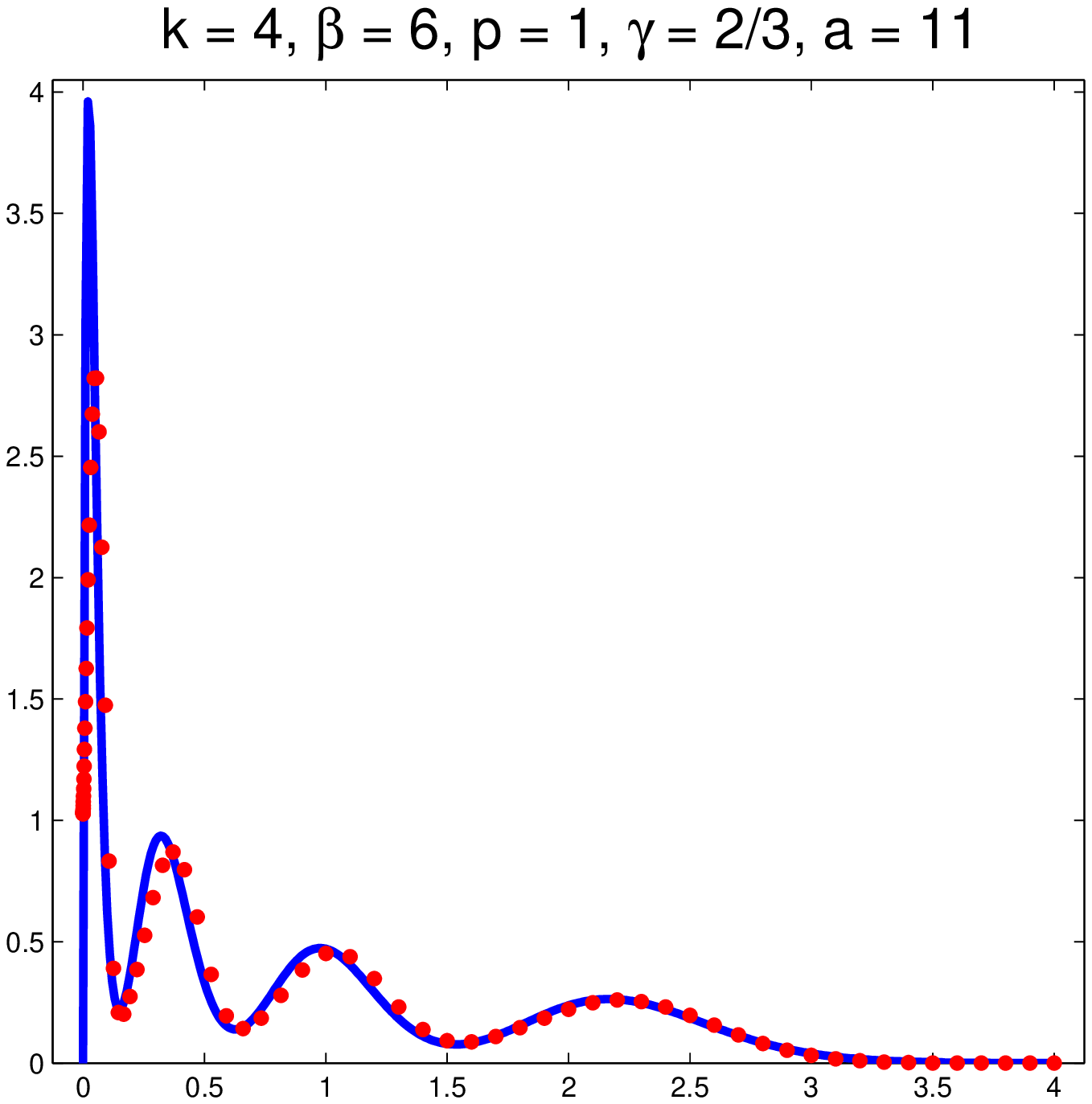, height = 6.9cm}} \\
\parbox[b]{7.1cm}{\epsfig{figure = 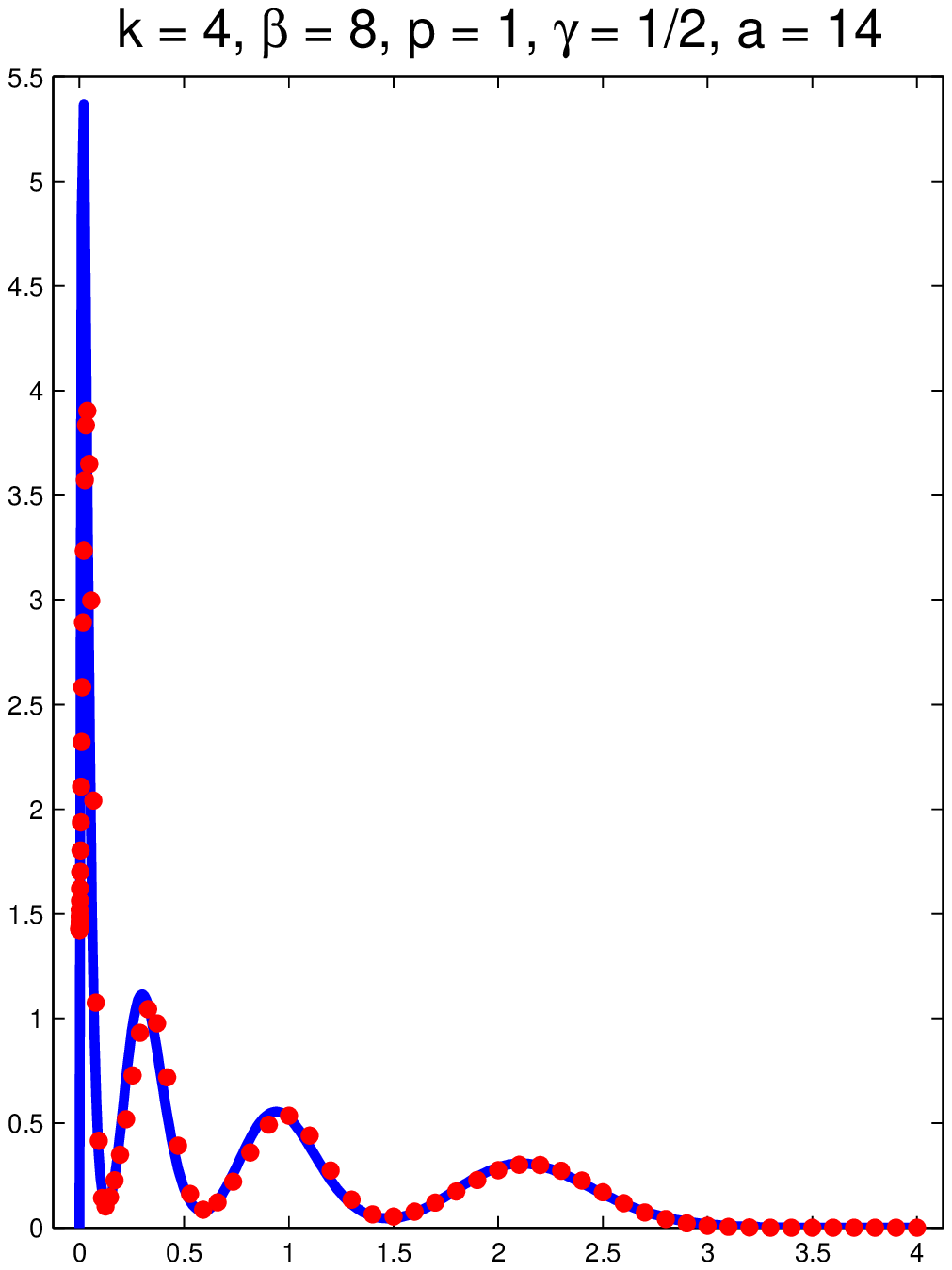, height = 8.9cm}} \hspace{.5cm}
\parbox[b]{7.1cm}{\epsfig{figure = 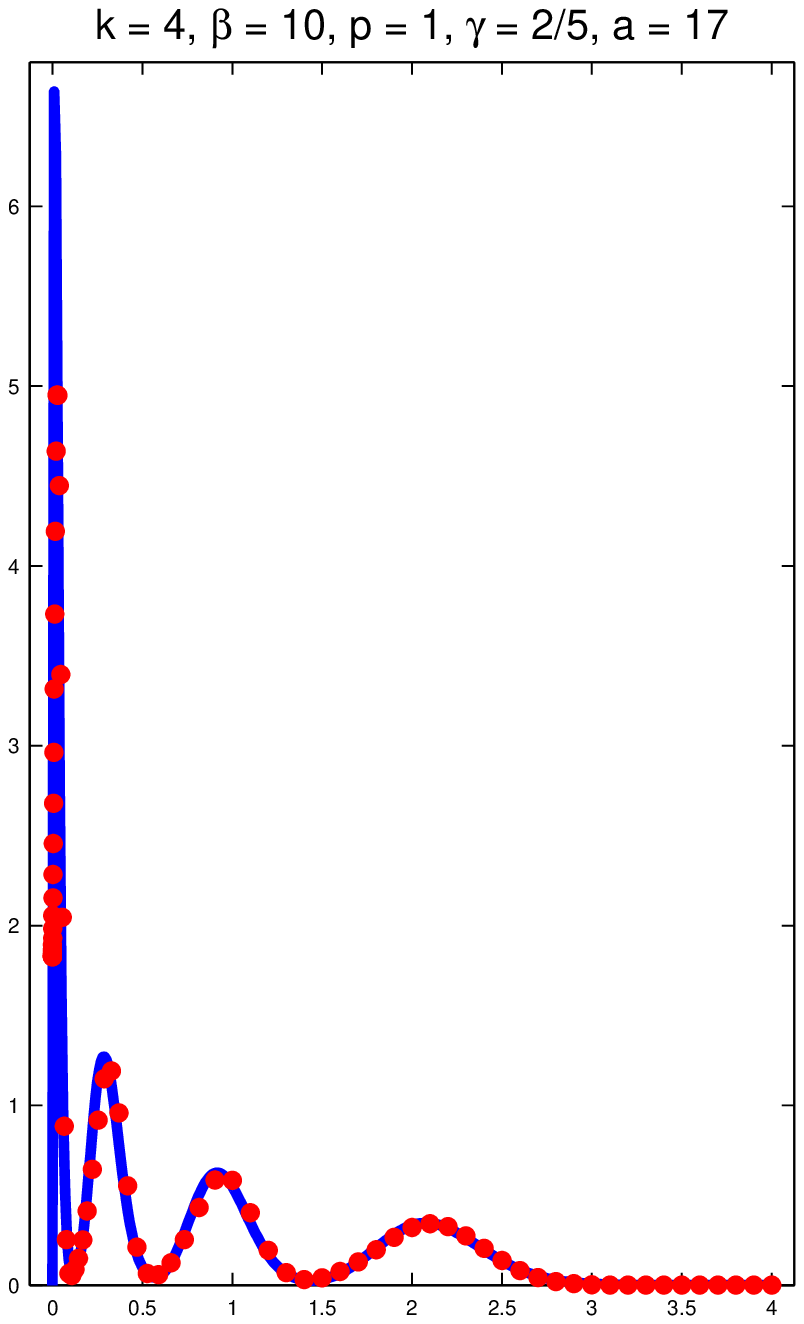, height = 10.85cm, width = 6.7cm}}
\caption{Laguerre case b):  sum of Gaussians approximation to the level densities (dots) and exact level densities (lines) for $k = 4$, $p=1$, and $\beta = 4,6,8,10$}\label{levdenslag_p1}
\end{figure}

The approximation works, once again, surprisingly well, as demonstrated by Figure \ref{levdenslag_p1}, where $n = 4$, $p=1$, $\beta = 4,6,8,10$, and $\gamma = 1, 2/3, 1/2, 2/5$ (or $a = 8, 11, 14, 17$). 

\begin{remark} Note that in this case, the smallest root converges to $0$ (which is the smallest root of the Laguerre polynomial $L^{-1}_4(x)$), and the presence of the delta function at $0$ in the sum of Gaussians is very clearly visible. \end{remark}

Thus we can conclude that in both cases, a good approximation is obtained even for $\beta$ relatively small.

\section{Circular Ensembles}

Similar to the $\beta$-Hermite ensemble, we have the circular ensembles defined by the joint eigenvalue $e^{i \theta_j}$ (with $\theta_j \in [0,1]^n$) density proportional to 
\begin{eqnarray*}
f_{\beta} \varpropto \prod_{1 \leq j <l \leq k} |e^{i \theta_j} - e^{i \theta_l}|^{\beta}~.
\end{eqnarray*}

The $\beta = 2$ circular ensemble is also known as the Haar measure on the unitary group $U_n$. The eigenvalues of $U_n$ appear to be almost uniformly distributed on the unit circle (see the experiment with $k = 100$ in Diaconis' paper \cite{diaconis03x}). For any fixed $k$, as $\beta \rightarrow \infty$, the eigenvalues freeze into place uniformly at the $k$th roots of unity. We believe that the same Gaussian phenomenon will hold, and the fluctuation of eigenvalue $i$ will behave like a normal centered at the $i$th root of the unity, with variance depending on $\frac{1}{\beta}$. 

\section{Acknowledgments}
The authors would like to thank the MIT-Singapore Alliance and the National Science Foundation for their support (NSF grant DMS9971591). Ioana Dumitriu's research was supported in part by a Miller Fellowship at U.C. Berkeley. She would like to thank Alexei Borodin for the useful comments and suggestions in presenting the results of this paper.

\bibliography{bib_10_01}
\bibliographystyle{plain}

\end{document}